\documentclass[11pt]{article}
\pdfoutput=1

\usepackage{amsmath}
\usepackage{amssymb}
\usepackage{amsfonts}
\usepackage{mathrsfs}

\usepackage{mathrsfs}
\usepackage{fullpage}
\usepackage{setspace}
\usepackage{bm}
\usepackage{bbm}
\usepackage{relsize}
\usepackage{adjustbox}
\usepackage{multirow}
\usepackage{cite}
\usepackage{filecontents}
\usepackage{scalerel}

\usepackage[normalem]{ulem}
\usepackage{enumerate}
\usepackage{yfonts}

\usepackage{psfrag}
\usepackage{array}
\usepackage{booktabs}
\newcolumntype{N}{>{\centering\arraybackslash}m{.5in}}
\newcolumntype{G}{>{\centering\arraybackslash}m{2in}}

\usepackage[latin1]{inputenc}
\usepackage{graphicx}
\usepackage{cancel}
\usepackage{slashed}
\usepackage{mathtools}

\usepackage[font=small,labelfont=bf]{caption}
\usepackage{subcaption}

\usepackage[colorlinks=true]{hyperref} 
\hypersetup{
    unicode=false,          % non-Latin characters 
    pdftoolbar=true,        % show Acrobat
    pdfmenubar=true,        % show Acrobat 
    pdffitwindow=false,     % window fit to page when opened
    pdfstartview={FitH},    % fits the width of the page to the window
    pdftitle={Exploring waveforms with non-GR deviations for extreme mass-ratio inspirals},    % title
    pdfauthor={Mostafizur Rahman, Shailesh Kumar and Arpan Bhattacharyya},     % author
    pdfnewwindow=true,      % links in new window
    colorlinks=true,       % false: boxed links; true: colored links
    linkcolor=blue,          % color of internal links (change box color with linkbordercolor)
    citecolor=red,        % color of links to bibliography
    filecolor=magenta,      % color of file links
    urlcolor=blue           % color of external links
    }
\usepackage{hyperref}
\hypersetup{colorlinks=true}

\def\equationautorefname~#1\null{%
	Eq.~(#1)\null
}
\def\figureautorefname~#1\null{%
	Fig.~#1\null
}
\def\tableautorefname~#1\null{%
	Table.~#1\null
}
\def\sectionautorefname~#1\null{%
	Section #1\null
}
\def\appendixautorefname~#1\null{%
	Appendix #1\null
}

\begin{document}
%%%%%%%%%%%%%%%%%%%%%%%%%%%%%%%%%%%%%%%%%%%%%%%%%%%%%%%%%%%%%%%%%%%%%%%%%%%%%%%%

%%%%%%%%%%%%%%%%%%%%%%%%%%%%%%%%%%%%%%%%%%%%%%%%%%%%%%%%%%%%%%%%%%%%%%%%%%%%%%%%
% Title page
%%%%%%%%%%%%%%%%%%%%%%%%%%%%%%%%%%%%%%%%%%%%%%%%%%%%%%%%%%%%%%%%%%%%%%%%%%%%%%%%
\numberwithin{equation}{section}
{
\begin{titlepage}
\begin{center}

\hfill \\
\hfill \\
\vskip 0.75in
%Probing dark matter signatures via gravitational wave fluxes of extreme mass-ratio inspiral
%Measuring gravitational wave fluxes with dark matter dynamical friction via extreme mass-ratio inspiral
{\Large {\bf Exploring waveforms with non-GR deviations for \\
extreme mass-ratio inspirals}}

%{\Large \bf Probing deformations with gravitational waves from \\
%extreme mass-ratio inspirals
%Gravitational waves from EMRIs as a probe of deformations to general relativity %Insights on detecting refined form of deformed-Kerr geometry with EMRIs %Prospects of detecting deviations to Kerr geometry with radiation reaction effects in EMRIs 
%What extreme mass-ratio inspiral can tell us about extra dimensions?
%Finding signature of higher dimensions through Extreme mass-ratio inspiral 
%Extreme mass-ratio inspiral: motion of a secondary around a brane black hole with a tidal charge}

\vskip 0.2in

{\large Shailesh Kumar${}$\footnote{\href{mailto: shailesh.k@iitgn.ac.in}{shailesh.k@iitgn.ac.in}}$^{a}$, Rishabh Kumar Singh${}$\footnote{\href{mailto: singhrishabh@iitgn.ac.in}{singhrishabh@iitgn.ac.in}}$^{a}$, Abhishek Chowdhuri${}$\footnote{\href{mailto: chowdhuri_abhishek@iitgn.ac.in}{chowdhuri\_abhishek@iitgn.ac.in}}}$^{a,b,c}$,\\ Arpan Bhattacharyya${}$\footnote{\href{mailto:  abhattacharyya@iitgn.ac.in}{abhattacharyya@iitgn.ac.in}}$^{a}$

\vskip 0.2in

{\it ${}$$^a$Indian Institute of Technology, Gandhinagar, Gujarat-382355, India\\ ${}$$^b$ Institute of Mathematical Sciences, CIT Campus, Tharamani, Chennai 600113, India\\${}$$^c$ Homi Bhabha National Institute, Training School Complex, Anushakti Nagar,\\ Mumbai 400085, India}

\vskip.5mm

\end{center}

\vskip 0.2in

\begin{center} 
{\bf ABSTRACT }
\end{center}
The fundamental process of detecting and examining the polarization modes of gravitational waves plays a pivotal role in enhancing our grasp on the precise mechanisms behind their generation. A thorough investigation is essential for delving deeper into the essence of gravitational waves and rigorously evaluating and validating the range of modified gravity theories. In this line of interest, a general description of black holes in theories beyond general relativity can serve a meaningful purpose where distinct deviation parameters can be mapped to solutions representing distinct theories. Employing a refined version of the deformed Kerr geometry, which is free from pathological behaviours such as unphysical divergences in the metric, we explore an extreme mass-ratio inspiral system, wherein a stellar-mass object perturbs a supermassive black hole. We compute the effects of deformation parameters on the rate of change of orbital energy and angular momentum, orbital evolution and phase dynamics with leading order post-Newtonian corrections. With the waveform analysis, we assess the plausibility of detecting deviations from general relativity through observations facilitated by the Laser Interferometer Space Antenna (LISA), simultaneously constraining the extent of these deviations. Therefore, this analysis provides an understanding while highlighting the essential role of observations in advancing gravitational phenomena beyond general relativity. 
\vfill

%\noindent \today

\end{titlepage}
}

%%%%%%%%%%%%%%%%%%%%%%%%%%%%%%%%%%%%%%%%%%%%%%%%%%%%%%%%%%%%%%%%%%%%%%%%%%%%%%%%
% Table of contents
%%%%%%%%%%%%%%%%%%%%%%%%%%%%%%%%%%%%%%%%%%%%%%%%%%%%%%%%%%%%%%%%%%%%%%%%%%%%%%%%
\newpage
\tableofcontents

\section{Introduction}
%Gravitational wave detections have been proven to uncover many insightful informations about the fundamental ideas of the laws governing gravity and also on gravitating systems
Gravitational wave (GW) detections have been proven to uncover insightful information about the fundamental laws governing gravity, enriching our cosmic understanding and testing boundaries beyond general relativity \cite{LIGOScientific:2016aoc, LIGOScientific:2016sjg, LIGOScientific:2017bnn}. These observations have also been crucial in providing new information regarding the population of binary black hole systems and have allowed us to perform novel tests of gravity in highly dynamical regimes \cite{KAGRA:2021vkt, LIGOScientific:2021sio}. However, with the planned next-generation ground-based and space-based detectors, the precision frontiers for GW observations and also the possibility of unveiling new physics will be pushed to unprecedented limits \cite{PhysRevD.95.103012, Berry:2019wgg, PhysRevD.103.044024, LISA:2022kgy}. It has also been speculated that GW observations can reveal information about the environment in which binary black hole systems are located \cite{PhysRevLett.107.171103, Zwick:2022dih, PhysRevLett.129.241103, CanevaSantoro:2023aol}. Given these numerous prospects, intermediate and extreme mass-ratio inspirals (IMRIs and EMRIs, respectively) have gained significant attention for being interesting sources \footnote{Interested readers are referred to \cite{Cardenas-Avendano:2024mqp} for a comprehensive review on this subject.}. They are expected to be observed with the upcoming space-based detector Laser Interferometer Space Antenna (LISA) \cite{PhysRevD.95.103012} and possibly with next-generation ground-based detectors in the case of IMRIs \cite{PhysRevD.98.063018}. To describe to the reader what these I/EMRI systems are, these are binary systems with highly asymmetric component masses typically ranging between $\sim 10^{-2}-10^{-4}$ for IMRIs and smaller than $\sim 10^{-4}$ for EMRIs \cite{PhysRevD.95.103012, PhysRevD.98.063018, Amaro-Seoane_2007}. Such systems typically complete a large number of GW cycles in the band, and hence, they become ideal probes to perform high-precision tests of gravity and also the environment surrounding astrophysical black holes \cite{PhysRevD.95.103012, Berry:2019wgg, PhysRevD.89.104059, PhysRevLett.129.241103, Destounis:2022obl,Rahman:2022fay,  Zi:2024dpi, AbhishekChowdhuri:2023cle, Rahman:2023sof, Datta:2024vll, Rahman:2021eay}.
\par 
Observations by the LIGO-Virgo-KAGRA collaborations (LVK) have indicated so far that there are not so much deviations from general relativity (GR) \cite{LIGOScientific:2016lio, LIGOScientific:2018dkp}. However, these provide cues to put strong constraints on any deviations from Einstein's theory. To fully utilize this new perspective on our universe, we must understand how deviations from general relativity (GR) arise and their effects on the gravitational wave (GW) waveform. The interest in modified gravity theories is driven by both observational insights and substantial theoretical advancements \cite{Sullivan:2019vyi, Fujii:2003pa}. Examples of the former are the desire to compare the increasing number of strong-field measurements with a sufficiently large class of consistent parametrized theoretical descriptions of black holes \cite{Collins:2004ex, Glampedakis:2005cf, Psaltis:2008bb, PhysRevD.87.124017}. Recent studies validate parametrized tests by confirming their consistency with general relativity. These studies also evaluate how well these tests can measure deviations, with possible constraints,  from general relativity with increasing GW observations \cite{Cornish:2011ys,Li:2011cg, Meidam:2017dgf, Tahura:2018zuq, Carson:2019yxq, Saleem:2021nsb, LIGOScientific:2021sio, Gupta:2020lxa}.
\par 
The new degrees of freedom (DOF) in modified gravity theories are crucial for showcasing intricate physics near black holes and neutron stars, which can be explored by analysing gravitational waves (GWs) from compact binary mergers. In scenarios with a scalar degree of freedom, the binary systems emit scalar radiations in addition to the tensorial radiation during the merging process. Since the scalar radiation emitted during the inspiral phase of binaries also modifies the gravitational waveform, it is possible to derive independent constraints on the scalar charge and model parameters of theories. In this vein, the gravitational waveforms in nonminimally coupled scalar-tensor theories have been computed in \cite{Lang:2013fna, Sennett:2016klh, Bernard:2018hta, Higashino:2022izi}. From an observational standpoint, there are compelling reasons to consider theories such as Hordenski's and beyond \cite{Horndeski:1974wa, AbhishekChowdhuri:2022ora, Higashino:2022izi, Dong:2022cvf, BenAchour:2024zzk, Toniato:2024gtx, Gao:2024rel, DeFalco:2024ojf, Battista:2021rlh, Battista:2022hmv}. These theories have successfully met rigorous constraints based on multi-messenger GW astronomy. The detection of the GW170817 binary neutron star merger, along with its associated electromagnetic counterparts, has provided an accurate bound on the speed of GWs \cite{Ezquiaga:2017ekz, Creminelli:2017sry}. The waveforms in which one is mainly interested can be constructed from various techniques, including the effective one-body (EOB) formalism \cite{Buonanno:1998gg, Buonanno:2000ef} and several perturbative methods for the inspiral phase, including the post-Newtonian (PN) \cite{Blanchet:2013haa, Lang:2013fna, Sennett:2016klh, Sagunski:2017nzb, Bernard:2018hta, Liu:2020moh, Higashino:2022izi, AbhishekChowdhuri:2022ora, Dong:2022cvf, BenAchour:2024zzk, Toniato:2024gtx, Zi:2023qfk} and post-Minkowskian (PM) approximations \cite{Damour:2016gwp, Damour:2017zjx}, as well as effective field theory (EFT) formalisms \cite{Porto:2016pyg, Schafer:2018kuf, Barack:2018yly, Levi:2018nxp, Bhattacharyya:2023kbh, Bhattacharyya:2024aeq}.
\par 
%In the inspiral phase of the binary, characterized by the weak-field regime, employing the post-Newtonian (PN) scheme to comprehend the dynamics of the binaries yields significant and non-trivial consequences.

%Studying systems like EMRIs, which have a highly asymmetric mass-ratio lying in the range ($q\equiv\mu/M = 10^{-7}-10^{-4}$), makes up for an effective picture where a smaller secondary (which is one of the components of the binary with mass $\mu$) moves around in geodesics cast by the geometry of the central supermassive black hole, called the primary (which is another component of the binary with mass $M$)
In the inspiral phase of a binary system, characterized by the weak-field regime, the post-Newtonian (PN) scheme provides valuable insights into the dynamics of the binaries \cite{Blanchet:2013haa}. This method leads to significant and nontrivial consequences for understanding their behaviour. Studying systems like EMRIs, whose highly asymmetric mass-ratio falls within the range ($q\equiv\mu/M = 10^{-7}-10^{-4}$), provides an interesting picture to examine strong gravity regimes. In such a picture, a smaller secondary object, one of the binary's components with mass $\mu$, moves in a geodesics cast by the geometry of the central supermassive black hole, called the primary object (another component of the binary with mass $M$). Several model-independent tests of the no-hair theorem have been suggested using GW observations of EMRIs \cite{Ryan:1995xi, PhysRevD.77.064022, PhysRevD.69.082005, PhysRevD.84.064016, Cardenas-Avendano:2024mqp, Gair:2011ym,Carson:2020iik}. Hence, exploring waveforms in this context can be a useful tool. 
 \par 
As part of a test of the ``no-hair" theorem, an obvious candidate to explore is deformed Kerr geometries, with these deformations being parametric. They have a smooth Kerr limit when the deviations vanish. Any deviation from the Kerr metric can lead to several interpretations. For example, within general relativity, such a deviation might indicate that the object is not a black hole instead it could be a stable stellar or an exotic object \cite{Collins:2004ex, PhysRevD.87.124017}.
%Any deviation from the Kerr metric implies some possible interpretations: one of them being within GR, the object, described by this kind of geometry, cannot be a black hole but instead can be a stable stellar or an exotic object \cite{Collins:2004ex, PhysRevD.87.124017}. 
 \par 
However, such deformations have been studied and have shown some pathologies as well \cite{PhysRevD.87.124017}. In \cite{PhysRevD.87.124017}, authors analyze such parametric deviations in detail and identify regions where these parameters are unphysical, and also comment on the range of coordinates, which gives us a possible way to get away from these pathologies. In follow-up works \cite{Carson:2020dez, Yagi:2023eap}, the authors propose a more general class of deformed rotating metrics and demonstrate by an example how a simple redefinition can handle divergences in the metric. In this paper, we take the cue of this metric and extend our previous study in the context of EMRIs \cite{AbhishekChowdhuri:2023gvu} to explore the non-trivial effects of these deviation parameters on the GW waveform. To explain this, we organize our paper in the following way: After a brief introduction to the refined deformed geometries in Section (\ref{BH}), we discuss the geodesic motions of the particle in such geometry in Section (\ref{gdscvecmtn}). This gives us an idea of the orbital dynamics of the inspiralling object. Next, in Section (\ref{radiation reaction}), we explore the radiation reaction effects on the system and compute the effects of deviations on orbital evolution and phase. Section (\ref{detect}) discusses the prospects of detectability with the waveform and calculating the mismatch. We conclude by discussing our results in Section (\ref{dscn}) with possible future directions.
 \par 
 \textit{Notation and Convention: } We set the fundamental constants $G$ and $c$ to unity and  adopt positive sign convention $(-1,1,1,1)$. Roman letters are used to denote spatial indices, and Greek letters are used to represent four-dimensional indices.

\section{Deformed Kerr-Geometry: A refined Parameterized Metric} \label{BH}
This section briefly discusses the refined version of a general stationary, axisymmetric, asymptotically flat and separable metric \cite{Yagi:2023eap}, which redefines the arbitrary radial functions that provide the remedy for the pathological behaviour in the original Johannsen spacetime \cite{PhysRevD.88.044002} and Carson-Yagi (CY) metric \cite{Carson:2020dez}. Such a metric is obtained by amending and employing the comprehensive study carried out by Johannsen and CY \cite{PhysRevD.88.044002, Carson:2020dez}. The general line element of the new refined metric is given by
\begin{align}\label{mets}
ds^{2} =& -\frac{\tilde{\Sigma}A_{5}}{\rho^{4}}(A_{5}-a^{2}A_{2}^{2}\sin^{2}\theta)dt^{2}+\frac{\tilde{\Sigma}}{A_{5}}dr^{2}+\tilde{\Sigma}d\theta^{2}+\frac{2aA_{5}\tilde{\Sigma}}{\rho^{4}}(A_{5}-A_{0})\sin^{2}\theta dtd\phi + \\ \nonumber 
& \frac{\tilde{\Sigma}A_{5}\sin^{2}\theta}{\rho^{4}}(A_{1}^{2}-a^{2}A_{5}\sin^{2}\theta)d\phi^{2}, 
\end{align}
where $\rho^{4} = a^{4}A_{2}^{2}A_{5}\sin^{4}\theta+a^{2}\sin^{2}\theta (A_{0}^{2}-2A_{0}A_{5}-A_{1}^{2}A_{2}^{2})+A_{1}^{2}A_{5}$, with further,
\begin{align}
\Tilde{\Sigma} \equiv \Sigma + f(r) + g(\theta) \hspace{3mm} ; \hspace{3mm} \Sigma \equiv r^{2}+a^{2}\cos^{2}\theta \hspace{3mm} ; \hspace{3mm} \Delta \equiv r^{2}+a^{2}-2Mr.
\end{align}
One can expand the arbitrary deviation functions around the spatial infinity in the following manner
\begin{equation}\label{sdfd}
\begin{aligned}
A_{i}(r) \equiv & \hspace{1mm}r^{2}\Big(1+\frac{a^{2}}{r^{2}}+\mathlarger{\mathlarger{\mathlarger{\Sigma}}}_{n=1}^{\infty}\alpha_{in}\Big(\frac{M}{r}\Big)^{n}\Big) \hspace{3mm} ; \hspace{3mm} A_{5} \equiv \hspace{1mm} r^{2}\Big(1-\frac{2M}{r}+\frac{a^{2}}{r^{2}}+\mathlarger{\mathlarger{\mathlarger{\Sigma}}}_{n=1}^{\infty}\alpha_{5n}\Big(\frac{M}{r}\Big)^{n}\Big)\,,\\
A_{2} \equiv & \hspace{1mm} 1+\mathlarger{\mathlarger{\mathlarger{\Sigma}}}_{n=1}^{\infty} \alpha_{2n}\Big(\frac{M}{r}\Big)^{n} \hspace{2mm} ; \hspace{2mm}
f(r) \equiv \hspace{1mm} r^{2}\mathlarger{\mathlarger{\mathlarger{\Sigma}}}_{n=1}^{\infty}\epsilon_{n}\Big(\frac{M}{r}\Big)^{n} \hspace{2mm} ; \hspace{2mm}
g(\theta) \equiv \hspace{1mm} M^{2}\mathlarger{\mathlarger{\mathlarger{\Sigma}}}_{n=0}^{\infty}\gamma_{n}P_{n}(\cos\theta),
\end{aligned}
\end{equation}
where $i\in (0, 1)$. $P_{n}(\cos\theta)$ denotes the Legendre polynomials, and the parameters ($\alpha_{in}, \epsilon_{n}$) describe the deviations from the Kerr geometry. We can expand the deformation/deviation functions in powers of $\frac{M}{r}$ with coefficients $\alpha_{in}$. We consider $g(\theta)=0$ for computational convenience without loss of generality \cite{Carson:2020dez}. This new form of the metric contains a new radial function $A_{0}$ that approaches the rescaled version of the Johannsen spacetime if taken $A_{0}\longrightarrow A_{1}A_{2}$ \cite{PhysRevD.88.044002}. Interestingly, the metric still shows the separable equations and brings forth a Carter-like constant, even though the deviations depend on ($r, \theta$). The other details of the metric have been discussed nicely in \cite{Yagi:2023eap}; however, we briefly touch upon the separability of geodesic motion in the appendix (\ref{apenteu1}), which will set up the basic ingredient to estimate the rate of change of orbital energy and angular momentum, and related quantities. Further, we expand the deviation functions in the following manner:

\begin{equation}\label{dfgn}
\begin{aligned}
A_{0} =& r^{2}\Big(1+\frac{a^{2}}{r^{2}}+\alpha_{01} \Big(\frac{M}{r}\Big)+\alpha_{02}\Big(\frac{M}{r}\Big)^{2}\Big)\,, \\
A_{1} =& r^{2}\Big(1+\frac{a^{2}}{r^{2}}+\alpha_{11} \Big(\frac{M}{r}\Big)+\alpha_{12}\Big(\frac{M}{r}\Big)^{2}+\alpha_{13}\Big(\frac{M}{r}\Big)^{3}\Big)\,, \\
A_{2} =& 1+\alpha_{21}\Big(\frac{M}{r}\Big)+\alpha_{22}\Big(\frac{M}{r}\Big)^{2} \,,\\
A_{5} =& r^{2}\Big(1-\frac{2M}{r}+\frac{a^{2}}{r^{2}}+\alpha_{51}\Big(\frac{M}{r}\Big)+\alpha_{52}\Big(\frac{M}{r}\Big)^{2}\Big)\,, \\
f(r) =& r^{2}\Big(\epsilon_{1}\Big(\frac{M}{r}\Big)+\epsilon_{2}\Big(\frac{M}{r}\Big)^{2}+\epsilon_{3}\Big(\frac{M}{r}\Big)^{3}\Big)\,,
\end{aligned}
\end{equation}
where ($\alpha_{01}, \alpha_{02}, \alpha_{11}, \alpha_{12}, \alpha_{13}, \alpha_{21}, \alpha_{22}, \alpha_{51}, \alpha_{52}, \epsilon_{1}, \epsilon_{2}, \epsilon_{3}$) are deformations to Kerr geometry. It is apparent from Eq. (\ref{sdfd}) that the contribution of deviations appears with the series expansion of $M/r$ in a perturbative manner. We have included the subleading terms of $M/r$ as well, as shown in Eq. (\ref{dfgn}). However, we examine only the leading order contribution of deviation parameters in the observable quantities and determine their detectability with LISA observations. In the present article, we use the refined metric Eq. (\ref{mets}) and refrain from imposing the parametrized-post-Newtonian (ppN) constraints because Birkhoff's theorem does not, in general, hold for non-GR theories \cite{Yagi:2023eap}. With such considerations, by performing the leading order PN analysis, we obtain the leading order contribution of deformations/deviations at the 0PN order itself, and it turns out that the parameters which will contribute at this order are ($\alpha_{11}, \alpha_{51}, \epsilon_{1}$). Moreover, the other deviation parameters generate a subleading effect. For instance, the contributions from ($\alpha_{12}, \epsilon_{2}$) are present at 1PN order. The terms ($\alpha_{01}, \alpha_{02}$) contribute at 1.5PN and 2PN orders, respectively, as well as the terms involving cross terms with spin $a$, represented as $\mathcal{O}(a\alpha_{01})$ and $\mathcal{O}(a\alpha_{02})$. Similarly, the terms ($\alpha_{12}, \alpha_{22}$) and the cross terms with spin $a$, i.e., $\mathcal{O}(a^{2}\alpha_{21})$ and $\mathcal{O}(a^{2}\alpha_{22})$ arise at higher PN orders arise at higher PN orders. As these effects are subleading, they do not affect the main findings of our analysis. On the other hand, with ppN constraints in \cite{PhysRevD.88.044002}, we have previously reported the emergence of deformation parameters ($\alpha_{13}, \alpha_{52}, \epsilon_{3}$) at 2PN order \cite{AbhishekChowdhuri:2023gvu}.

Following \cite{AbhishekChowdhuri:2023gvu}, in principle, to compute the waveform up to 2PN, one should be able to construct all relevant quantities that involve deformations contributing at all orders up to 2PN, including the GR results. However, this pertains to numerous significant advancements that will be addressed in future studies. For a detailed discussion, readers are encouraged to look into \cite{Canizares:2012is}, where a similar study has been carried out for Dynamical Chern Simons Modified Gravity theory (DCSMG). Since we are not focusing on any particular theory, the metric carries several deformation parameters that can be mapped to different spacetimes\textcolor{red}{\footnote{The reader is encouraged to look into \cite{Yagi:2023eap} for a comprehensive review of the mapping to other spacetime solutions for alternate theories.}}; thus, we investigate our analysis with leading order PN results without any ppN constraints.

\section{Eccentric orbital motion }\label{gdscvecmtn}
This section provides necessary expressions and details of the eccentric orbital dynamics of the inspiralling object that accompanies the deviation parameters present in Kerr-deformed geometry. In general, the Kerr spacetime exhibits the three constants of motion- energy ($E$), angular momentum ($J_{z}$) and the Carter constant ($\mathcal{Q}$). As we will consider equatorial orbits, that sets the Carter constant to be zero, $\mathcal{Q}=0$ \cite{Glampedakis:2002cb, Glampedakis:2002ya, PhysRevD.61.084004}. We perform our computation by making the quantities dimensionless with respect to the central black hole mass $M$ \cite{AbhishekChowdhuri:2023cle}; however, we avoid giving any special notation to such dimensionless quantities for writing convenience. One can always express the quantities in physical units at a later stage. In addition, geodesic velocities can be written in the following way, as described in (\ref{apenteu1}),

\begin{equation}
\begin{aligned}\label{gdscs1}
\frac{dt}{d\tau}  =& \frac{1}{A_{5} \Sigma}(A_{1}^2 E+a J_{z} (A_{5}-A_{0}))\,, \\
\frac{d\phi}{d\tau} =& \frac{1}{A_{5} \Sigma}(A_{5} J_{z} \csc ^2\theta-a \left(A_{5}-A_{0}\right)E)\,, \\
\Big(\frac{dr}{d\tau}\Big)^{2} =& \frac{A_{5}}{\Sigma^{2}}\Big[-C-\mu^{2}(f+r^{2})-\frac{1}{A_{5}}(-a^{2}J_{z}^{2}A_{2}^{2}+2aEJ_{z}A_{0}-E^{2}A_{1}^{2}) \Big]\,, \\
\Big(\frac{d\theta}{d\tau}\Big)^{2} =& \frac{1}{\Sigma^{2}}\Big[C-(a^{2}E^{2}\sin^{2}\theta+\mu^{2}a^{2}\cos^{2}\theta-2aEJ_{z}+J_{z}^{2}\csc^{2}\theta) \Big]\,,
\end{aligned}
\end{equation}
where $\mathcal{Q}\equiv C-(J_{z}-aE)^{2}$. Since $\mathcal{Q}=0$, it implies that $C=J_{z}^{2}-2aEJ_{z}+\mathcal{O}(a^{2})$, and all velocities have been computed up to $\mathcal{O}(a)$ as also discussed in the appendix (\ref{apenteu1}). We further examine the radial equation Eq. (\ref{gdscs1}) that provides the effective potential ($V_{eff}$), which captures the orbital motion of the inspiralling object. We can write down the $V_{eff}$ in terms of metric coefficients in the following way, 
\begin{align}\label{vef1}
V_{eff}(r) \equiv -\frac{1}{2}(g^{tt}E^{2}-2g^{t\phi}EJ_{z}+g^{\phi\phi}J_{z}^{2}+1)=\Big(\frac{dr}{d\tau}\Big)^{2}\,.
\end{align}
Since there are two turning points for the eccentric dynamics: \textit{periastron} ($r_{p}$) and \textit{apastron} ($r_{a}$), the bounded orbits can be described with the semi-latus rectum ($p$) and eccentricity ($e$): $r_{p}=p/(1+e)$ and $r_{a}=p/(1-e)$. Solutions for such orbits can be obtained in the range $r_{p}<r<r_{a}$ if $V_{eff}(r)<0\,.$ This is maintained only when $V'_{eff}(r_{a})>0$ and $V'_{eff}(r_{p})\leq 0$ \cite{Rahman:2023sof, AbhishekChowdhuri:2023gvu}, with prime denoting derivative with respect to the radial coordinate. As we know, the radial velocity vanishes at the turning points; hence $V_{eff}\vert_{r=r_{p},r_{a}}=0$. With the help of Eq. (\ref{vef1}) and turning points, we obtain the following positive roots of the constants of motion ($E, J_{z}$) of the object hovering in the vicinity of the deformed-Kerr geometry 
\begin{equation}
\begin{aligned}\label{cnst1}
E =& \sqrt{\frac{(p-2)^2-4 e^2}{p \left(p-3-e^2\right)}} + \frac{(e^{2}-1)}{p^{3/2}}\sqrt{\frac{(p-2)^{2}-4e^{2}}{(p-3-e^{2})^{3}}}\Big[\frac{(p-4)}{2}\alpha_{11}+\frac{(e^{2}-1)}{p}\alpha_{12} \Big] \\
& +\frac{(1-e^{2})(p^{2}-8p+4e^{2}+12)}{(p-3-e^{2})^{3/2}\sqrt{p((p-2)^{2}-4e^{2})}}\frac{\alpha_{51}}{4} +\frac{(1-e^{2})}{4}\sqrt{\frac{(p-2)^{2}-4e^{2}}{p^{3}(p-3-e^{2})}}\epsilon_{1}\,, \\
J_{z} =&\frac{p}{\sqrt{p-3-e^2}} + \frac{((p-2)^{2}-4e^{2})}{(p-3-e^{2})^{3/2}}\Big[\frac{\alpha_{11}}{2} + \frac{\alpha_{12}}{p}\Big] -\frac{1}{4\sqrt{p-3-e^{2}}}\Big[\frac{p^{2}\alpha_{51}}{(p-3-e^{2})}+(p-4)\epsilon_{1} \Big] \\
& -\frac{\sqrt{p-3-e^{2}}}{2p}\epsilon_{2}\,.
\end{aligned}
\end{equation}
As mentioned earlier, we have only considered the leading order corrections in deformation parameters. With the use of Eqs. (\ref{vef1}) and (\ref{cnst1}), we determine the location of the last stable orbit (LSO) that provides the region for ($p, e$), which separates the bound and unbound orbits \cite{Glampedakis:2002cb, PhysRevD.50.3816, PhysRevD.77.124050}. It also gives the lowest allowed value of the semi-latus rectum ($p_{sp}$) for all the bound orbits with a given $e$. The set of such points is termed \textit{separatrix}, which, for the spacetime under consideration, becomes the following:
\begin{equation}
\begin{aligned}\label{sp1}
p_{sp}=& 2(e+3)+\frac{8(1+e)}{(3+e)^{2}}\alpha_{12}-(e+3)\alpha_{51}+ \frac{(3+2e-e^{2})}{(3+e)}\epsilon_{1}\,.
\end{aligned}
\end{equation}
With the expression given in Eq. (\ref{sp1}), it is obvious to see that the separatrix reduces to $p_{sp}=6$ for the circular orbits in the Schwarzschild background; such orbits are called the innermost stable circular orbits (ISCOs); however, for the eccentric orbital motion, it gives $p_{sp}=6+2e$. The results comply with \cite{PhysRevD.50.3816, Glampedakis:2002ya}. 

We recall that the inspiralling object exhibits bounded motion in the region ($r_{p}, r_{a}$). It is useful to introduce a parametrization for the radial coordinate in order to overcome the diverging behaviour of differential equations at the turning points. The parametrization is 
% To put it another way, the separatrix determines the lowest semi-latus rectum value for which spacetime permits bound orbits for a given eccentricity. It is the last stable orbit of the motion. For circular orbits ($e=0$), it is called the innermost stable circular orbit (ISCO). The Eq.(\ref{sp1}) corresponds to the prograde motion of the inspiralling object and coincides with \cite{Glampedakis:2002ya} in leading order $a$; however, one can replace $a\rightarrow -a$ for the retrograde motion. It implies that the separatrix curve for prograde (retrograde) orbits will shift to the left (right) with respect to the Schwarzschild curve ($p^{sch}_{sp}=6+2e$) as the black hole spins up \cite{Glampedakis:2002ya}. 
% The motion occurs between $r_{p}$ to $r_{a}$ and vice-versa. As we choose $r$ as the parameter throughout the orbit, we can integrate Eqs.(\ref{gdscs2n}) by removing $\tau$ from the set of equations. We parametrize the radial coordinate $r$ in the following way to conquer the divergences at the turning points ($V_{eff}=0; r_{a}, r_{p}$)
\begin{align}\label{prmtz}
r=\frac{p}{1+e\cos\chi}\,.
\end{align}
These points ($r_{p}, r_{a}$) correspond to ($\chi=\pi, \chi=0$), respectively. In addition, eccentric motion exhibits two fundamental frequencies: radial ($\Omega_{r}$) and azimuthal ($\Omega_{\phi}$). Following \cite{PhysRevD.50.3816}, expressions for ($\Omega_{\phi}, \Omega_{r}$) are given by
\begin{equation}
\begin{aligned}\label{freq}
\Omega_{\phi} =& \frac{J_{z}}{Er^{3}}(r-2)-\frac{J_{z}(r-2)}{Er^{3}}\Big(-\alpha_{51}+\frac{2\alpha_{11}}{r}+\frac{2\alpha_{12}}{r^{2}}\Big)\,, \\
\Omega_{r} =& \frac{2\pi}{T_{r}} \hspace{3mm} ; \hspace{3mm} T_{r} = \int_{0}^{2\pi}d\chi\frac{dt}{d\chi} \hspace{3mm} ; \hspace{3mm} \frac{dt}{d\chi} = \frac{dt}{dr}\frac{dr}{d\chi}\,,
\end{aligned}
\end{equation}
where $T_{r}$ denotes the radial time period. We use these quantities further to compute the average rate of change of orbital energy and angular momentum, as well as phase, and examine the leading order contributions coming from these deviation parameters.

%%%%%%%%%%%%%%%%%%%%%%%%%%%%%%%%%%%%%%%%%%%%%%%%%%%%%%%%%%%%%%%%%%%%%%%%%%%%%%%%%%%%%%%%%%%%%%%%%%%%%%%%%%%%%%%%

\section{Fluxes, orbital evolution and phase} \label{radiation reaction}

In this section, we study the effect of radiation reaction on the inspiraling object, and, as a consequence, we estimate the rate of change of orbital energy and angular momentum, including the eccentric orbital dynamics that record the imprints of deformations. As the test body perturbs the geometry of the central supermassive black hole by inspiralling in its vicinity, such a motion generates GWs and brings the notion of radiation reaction effect, which allows the constants of motion associated with the moving object to evolve slowly in time. We consider such a scenario in this article and analyze the effects of the leading order deformations and mass-ratio \cite{Flanagan:2007tv, PhysRevD.52.R3159, Ryan:1995xi} with the equatorial consideration. Let us first introduce the basic ingredients to establish the base for flux computation.

We re-state analytical expressions of the constants of motion, derived in the appendix (\ref{apenteu1}), and represent them in the Cartesian coordinates ($x_{1},x_{2},x_{3})=(r\sin\theta \cos\phi, r\sin\theta \sin\phi, r\cos\theta$) that further helps us in estimating the rate change of constants of motion ($\mathcal{E}, J_{z}$). The required expressions are given as
\begin{equation}
\begin{aligned}\label{re}
\mathcal{E} =& \frac{1}{2} \dot{x}_{i}\dot{x}_{i}-\frac{1}{\sqrt{x_{i}x_{i}}}\Big(1-\frac{1}{2}(-2\alpha_{11}+\alpha_{51}+\epsilon_{1})\Big)+\frac{1}{2x_{i}x_{i}}(16\alpha_{11}-2\alpha_{12}-8\alpha_{51}-6\epsilon_{1}+\epsilon_{2}+8)\,,\\
J_{z} =& \epsilon_{3jk}x_{j}\dot{x}_{k}\Big(1+\frac{\epsilon_{2}}{x_{i}x_{i}}+\frac{\epsilon_{1}}{\sqrt{x_{i}x_{i}}}\Big),
% \mathcal{Q}+J_{z}^{2} =& (\epsilon_{ijk}x_{j}\dot{x}_{k})(\epsilon_{ilm}x_{l}\dot{x}_{m})-4a\frac{\epsilon_{3jk}x_{j}\dot{x}_{k}}{\sqrt{x_{i}x_{i}}},
% \Big(2+\frac{\alpha_{22}}{\sqrt{x_{i}x_{i}}}\Big)
\end{aligned}
\end{equation}
where we have used $r^{2}\sin^{2}\theta \dot{\phi}=\epsilon_{3jk}x_{j}\dot{x}_{k}$ and $r^{4}(\dot{\theta}^{2}+\sin^{2}\theta\dot{\phi}^{2})=(\epsilon_{ijk}x_{j}\dot{x}_{k})(\epsilon_{ilm}x_{l}\dot{x}_{m})$. The dot signifies differentiation with respect to the coordinate time $t$. These findings align with prior studies \cite{Flanagan:2007tv, AbhishekChowdhuri:2023gvu} with $\theta=\pi/2$. Note that we may anticipate several deformation parameters being considered at this stage. However, in the expressions for the averaged quantities, there are only three parameters that appear as the leading order corrections ($\alpha_{11}, \alpha_{51}, \epsilon_{1}$). Due to the radiation reaction effect, one can now write down the rate change of constants of motion, also called the instantaneous fluxes. In Eq. (\ref{re}), our interest lies in the first term of ($\mathcal{E}, J_{z}$) as we notice that only the first term has velocity dependence, eventually contributing as a radiation reaction acceleration while computing the instantaneous fluxes. Therefore, the rate change of constants of motion can be written in the following fashion,
\begin{align}\label{inst fluxes}
\dot{\mathcal{E}} = x_{i}\Ddot{x}_{i} \hspace{5mm} ; \hspace{5mm}
\dot{J}_{z} = \epsilon_{3jk}x_{j}\Ddot{x}_{k}\Big(1+\frac{\epsilon_{2}}{x_{i}x_{i}}+\frac{\epsilon_{1}}{\sqrt{x_{i}x_{i}}}\Big). 
%\hspace{2mm} ; \hspace{2mm}
% \dot{\mathcal{Q}+J_{z}^{2}} = 2(\epsilon_{ijk}x_{j}\dot{x}_{k})(\epsilon_{ilm}x_{l}\dot{x}_{m})-\frac{4a\epsilon_{3jk}x_{j}\Ddot{x}_{k}}{\sqrt{x_{i}x_{i}}}.
\end{align}
Hence, Eq. (\ref{inst fluxes}) has acceleration ($\Ddot{x}_{i}$) terms contributing to the radiation reaction, which is often denoted as $a_{j}$. It contains two symmetric trace-free (STF) contributions- mass ($I_{jk}$) and current ($J_{jk}$) quadrupole moments, which are given as
\begin{equation}\label{moments}
I_{jk}=\Big[x_j x_k\Big]^{\text{STF}} \hspace{3mm} ; \hspace{3mm}
J_{jk}=\Big[x_{j}\epsilon_{kpq}x_{p}\dot{x}_{q}-\dfrac{3}{2}x_jJ\delta_{k3}\Big]^{\text{STF}}.
\end{equation}
With this, the general expression of the radiation reaction acceleration, also denoted as $\Ddot{x}_{i}$, is given by \cite{Flanagan:2007tv, PhysRevD.52.R3159}
\begin{align}\label{accelration}
a_j=-\dfrac{2}{5}I^{(5)}_{jk}x_{k}+\dfrac{16}{45}\epsilon_{jpq}J^{(6)}_{pk}x_{q}x_{k}+\dfrac{32}{45}\epsilon_{jpq}J^{(5)}_{pk}x_{k} \dot{x}_{q}+\dfrac{32}{45}\epsilon_{pq[j}J^{(5)}_{k]p}x_{q} \dot{x}_{k}+\dfrac{8J}{15}J^{(5)}_{3i},
\end{align}
where $B_{[ij]}$ is an anti-symmetric quantity, written as $B_{[ij]}=\frac{1}{2}(B_{ij}-B_{ji})$. The superscripts denote the derivative order, and $J$, in the last term, refers to the black hole spin $a$. Finally, we use the set of Eqs. (\ref{moments}, \ref{accelration}) for computing the rate change of constants of motion represented in Eq. (\ref{inst fluxes}). We further utilize the expressions of velocities mentioned in Eq. (\ref{gdscs3n}), and as a result, with the leading order corrections, we obtain,
\begin{equation}
\begin{aligned}\label{enrjz2}
\mathcal{E} = \frac{(e^{2}-1)}{2p}\Big(1+\alpha_{11}-\frac{\alpha_{51}}{2}-\frac{\epsilon_1}{2}\Big) \hspace{3mm} ; \hspace{3mm}
J_{z} = \sqrt{p}\Big(1+\frac{\alpha_{11}}{2}-\frac{\alpha_{51}}{4}-\frac{\epsilon_{1}}{4}\Big)\,.
\end{aligned}
\end{equation}
This fully determines the instantaneous fluxes in terms of ($p, e, \chi$). Once we have ($\Dot{\mathcal{E}}, \Dot{J}_{z}$) following \cite{AbhishekChowdhuri:2023gvu}, our objective is to compute the average rate of change for these parameters. Adopting the adiabatic approximation \cite{Hinderer:2008dm, PhysRevD.103.104014, PhysRevLett.128.231101, Glampedakis:2002ya}, where the particle's motion can be approximated by geodesics, we are specifically concerned with temporal scales considerably shorter than those associated with the radiation reaction time scale. Hence, in this context, the system's radiative losses lead to slow changes in instantaneous fluxes. We estimate the mean (average) values of these quantities, denoted as ($\Dot{\mathcal{E}}, \Dot{J_{z}}$), by averaging over the course of a single orbit which can collectively be written as
\begin{align}
<\Dot{\mathcal{E}}> = \frac{1}{T_{r}}\int_{0}^{2\pi}\Dot{\mathcal{E}}\frac{dt}{d\chi}d\chi \hspace{3mm} ; \hspace{3mm} <\Dot{J_{z}}> = \frac{1}{T_{r}}\int_{0}^{2\pi}\Dot{J_{z}}\frac{dt}{d\chi}d\chi,
\end{align}
where,
\begin{align}
T_{r} = \frac{2\pi p^{3/2}}{(1-e^{2})^{3/2}}\Big(1-\frac{\alpha_{11}}{2}+\frac{\alpha_{51}}{4}+\frac{\epsilon_{1}}{4}\Big)\,.
\end{align}
As a result, the average loss of energy and angular momentum is
\begin{equation}
\begin{aligned}\label{avflxn}
<\Dot{\mathcal{E}}> =& -\frac{(1-e^{2})^{3/2}}{30p^{5}} (96+292e^{2}+37e^{4})(2+6\alpha_{11}-3\alpha_{51}-3\epsilon_{1})\,, \\
<\Dot{J}_{z}> =& -\frac{(1-e^{2})^{3/2}}{5p^{7/2}}(8+7e^2)(4+10\alpha_{11}-5\alpha_{51}-5\epsilon_{1}).
\end{aligned}
\end{equation}
The obtained averaged energy and angular momentum fluxes, if the deviation parameters are switched off, are consistent with the standard results present in the literature \cite{AbhishekChowdhuri:2023gvu, Flanagan:2007tv, Glampedakis:2002ya, Ryan:1995xi, PhysRev.131.435, PhysRev.136.B1224}. In order to verify the PN order, we can restore the power of the speed of light $c$ that implies $<\Dot{\mathcal{E}>}\sim -\frac{1}{p^{5}c^{5}}a_{1}(2+6\alpha_{11}-3\alpha_{51}-3\epsilon_{1})$ and $<\Dot{J}_{z}>\sim -\frac{1}{p^{7/2}c^{5}}a_{2}(4+10\alpha_{11}-5\alpha_{51}-5\epsilon_{1})$. Where ($a_{1}, a_{2}$) are functions of eccentricity, contained in Eq. (\ref{avflxn}), given by $a_{1}=\frac{1}{30}(1-e^{2})^{3/2}(96+292e^{2}+37e^{4})$ and $a_{2}=\frac{1}{5}(1-e^{2})^{3/2}(8+7e^2)$. As we take the adiabatic approximation, we can write down the balance equations in the following way\footnote{\textcolor{black}{It is to note that, to compute the GW flux, one needs to add the metric perturbation to this deformed Kerr background $\sim g_{\mu \nu,\text{dKerr}}+h_{\mu \nu}$, where $g_{\mu \nu,\text{dKerr}}$ is the background and $h_{\mu \nu}$ is the GW perturbation. In this paper, we are only considering the correction to the rate of change of orbital energy and angular momentum coming from the deformation parameters. It is not obvious a priori that the correction to GW flux from these deformation parameters would be subleading compared to the correction to the orbital energy and angular momentum. Computing such corrections to GW flux is beyond the scope of this paper, and we leave it for future studies, instead we work with approximated balance law as mentioned in (\ref{blnceqn}).}}:
\begin{align}\label{blnceqn}
\Big\langle\frac{d\mathcal{E}}{dt}\Big\rangle_{GW} = -\Big\langle\frac{d\mathcal{E}}{dt}\Big\rangle \hspace{5mm} ; \hspace{5mm} \Big\langle\frac{dJ_{z}}{dt}\Big\rangle_{GW} = -\Big\langle\frac{dJ_{z}}{dt}\Big\rangle\,.
\end{align}
it suggests that the average radiated energy and angular momentum flux correspond to the loss in orbital energy and angular momentum. Thus the expressions provide us the leading order corrections to the zeroth order PN results of GW fluxes, indicating the contributions from the theories beyond general relativity.

Next we compute the orbital evolution of the inspiralling object. Using Eq.(\ref{avflxn}), we have: $\frac{d\mathcal{E}}{dt} = \frac{\partial \mathcal{E}}{\partial p}\frac{dp}{dt} + \frac{\partial\mathcal{E}}{\partial e}\frac{de}{dt}$ and $\frac{dJ_{z}}{dt} = \frac{\partial J_{z}}{\partial p}\frac{dp}{dt} + \frac{\partial J_{z}}{\partial e}\frac{de}{dt}$ which further results in average rate change of ($p, e$),
\begin{equation}
\begin{aligned}
\Big\langle\frac{dp}{dt}\Big\rangle = \Big(\frac{\dot{\mathcal{E}}\partial_{e}J_{z} -\dot{J}_{z}\partial_{e}\mathcal{E}}{\partial_{p}\mathcal{E}\partial_{e}J_{z}-\partial_{e}\mathcal{E}\partial_{p}J_{z}}\Big) \hspace{3mm} ; \hspace{3mm} \Big\langle\frac{de}{dt}\Big\rangle = \Big(\frac{\dot{J}_{z}\partial_{p}\mathcal{E}-\dot{\mathcal{E}}\partial_{p}J_{z}}{\partial_{p}\mathcal{E}\partial_{e}J_{z}-\partial_{e}\mathcal{E}\partial_{p}J_{z}}\Big)\,.
\end{aligned}
\end{equation}
The ultimate expressions take the following form,
\begin{equation}
\begin{aligned}\label{dpdtn}
\Big\langle\frac{dp}{dt}\Big\rangle =& -\frac{8(1-e^2)^{3/2}}{5p^{3}}(8+7e^{2})(1+2\alpha_{11}-\alpha_{51}-\epsilon_{1})\,, \\
\Big\langle\frac{de}{dt}\Big\rangle =& -e\frac{(1-e^2)^{3/2}}{15p^4}(304+121e^{2}) (1+2\alpha_{11}-\alpha_{51}-\epsilon_{1})\,.
\end{aligned}
\end{equation}
The expressions in Eq. (\ref{dpdtn}) imply that, as a consequence of gravitational radiation, the semi-latus rectum ($p$) and eccentricity ($e$) gradually decrease over time. This decrease persists as long as the contributions from the terms proportional to the deformation parameters are subleading. The primary reason for this behaviour is that the leading terms on the right-hand sides of (\ref{dpdtn}) are negative, which ensures a continuous decrease in both ($p, e$) over time. Since the behaviour of orbital evolution is similar to \cite{AbhishekChowdhuri:2023gvu} except for the changes in values of deviation parameters; so we do not explicitly provide plots for this case. Interestingly, we notice that $dp/de$ is unaffected by the deviations. From Eq. (\ref{dpdtn}), we obtain
\begin{align} \label{4.12}
p = c_{0} e^{12/19} \Big(1+\frac{121}{304}e^{2}\Big)^{870/2299},
\end{align}
where $c_{0}$ completely depends on the initial conditions ($p_{0}, e_{0}$). This is the standard result established in the literature, which coincides with \cite{PhysRev.136.B1224, Maggiore:2007ulw}. The nonappearance of the deviation parameters in  Eq. \eqref{4.12} was also observed in \cite{AbhishekChowdhuri:2022ora} \footnote{This is possibly due to the fact that we are only considering the leading correction coming from the gravitational part, neglecting the effect of other degrees of freedom, which must be present in the theory to support these extra deviation parameters, e.g in \cite{AbhishekChowdhuri:2022ora} it was shown that for Horndeski theory, if one also add the contribution coming from scalar flux, the correction terms appear in $dp/de\,.$ Same thing is observed later in \cite{Trestini:2024zpi} for Scalar-Tensor theory well.}.

Further analysis can determine the time required for the inspiralling object to reach the LSO. We utilize Eq. (\ref{dpdtn}) and integrate it from $p=14$ to $p=p_{sp}$. In this analysis, we specifically focus on how the presence of deviation parameters affects this timescale, denoted as $\Delta t$, compared to the GR results. This method offers a means of assessing the effects of these parameters on the inspiral process. Therefore, by integrating the first expression in Eq. (\ref{dpdtn}) to determine the timescale, we proceed by subtracting the component contributing from GR. This process permits us to isolate the effects of the deviation parameters. Consequently, we arrive at:
\begin{align}\label{DeltaT}
\Delta t \approx \frac{5}{32}\frac{\left(1-e^2\right)^{-3/2}}{\left(7 e^2+8\right)}\left(p_{sp}^4-38416\right)(2 \alpha_{11}-\alpha_{51}-\epsilon_{1}),
\end{align}
where $p_{sp}$ denotes the truncation point of the inspiral, i.e., LSO; while using Eq. (\ref{sp1}), one should only consider taking the GR results for $p_{sp}$ since we are examining the leading order effects of deviation parameters which are already there in the Eq. (\ref{DeltaT}). This examination elucidates the change in the timescale required for the inspiralling object to attain the LSO, which is ultimately tied to the relative signs of the deviation parameters, underscoring their pivotal role in shaping the dynamics of the inspiral process.

%\subsection{Orbital phase} \label{dephasing1}
In Sec. (\ref{gdscvecmtn}), it was discussed that the eccentric motion within the system is characterized by two primary frequencies: $\Omega_{\phi}$ for angular motion and $\Omega_{r}$ for radial motion. Both of these frequencies play roles in phase computation. When considered together, we get the following:
\begin{align}
\frac{d\varphi_{i}}{dt} = \langle \Omega_{i}(p(t),e(t))\rangle = \frac{1}{T_{r}}\int_{0}^{2\pi}d\chi\frac{dt}{d\chi} \Omega_{i}(p(t),e(t),\chi) \hspace{5mm} ; \hspace{5mm} i = (\phi, r),
\end{align}
where $\langle\Omega_{i}\rangle$ denotes the averaged orbital frequency. The expression completely depends on ($p(t), e(t)$) apart from deviation parameters. We assume that the azimuthal frequency primarily regulates the analysis: $\varphi_{\phi}(t)\sim\phi (t)$ with $\varphi_{i}(0)=0$. Using Eq. (\ref{gdscs3n}), we obtain
\begin{align}\label{phase}
\frac{d\phi}{dt} = \Big(\frac{1-e^2}{p}\Big)^{3/2}\Big(1+\frac{\alpha_{11}}{2}-\frac{\alpha_{51}}{4}-\frac{\epsilon_{1}}{4}\Big)\,.
\end{align}
\begin{figure}[h!]
	%%%%%%%%%%%%%%%%%%%%%%%%
	\centering
	\minipage{0.33\textwidth}
	\includegraphics[width=\linewidth]{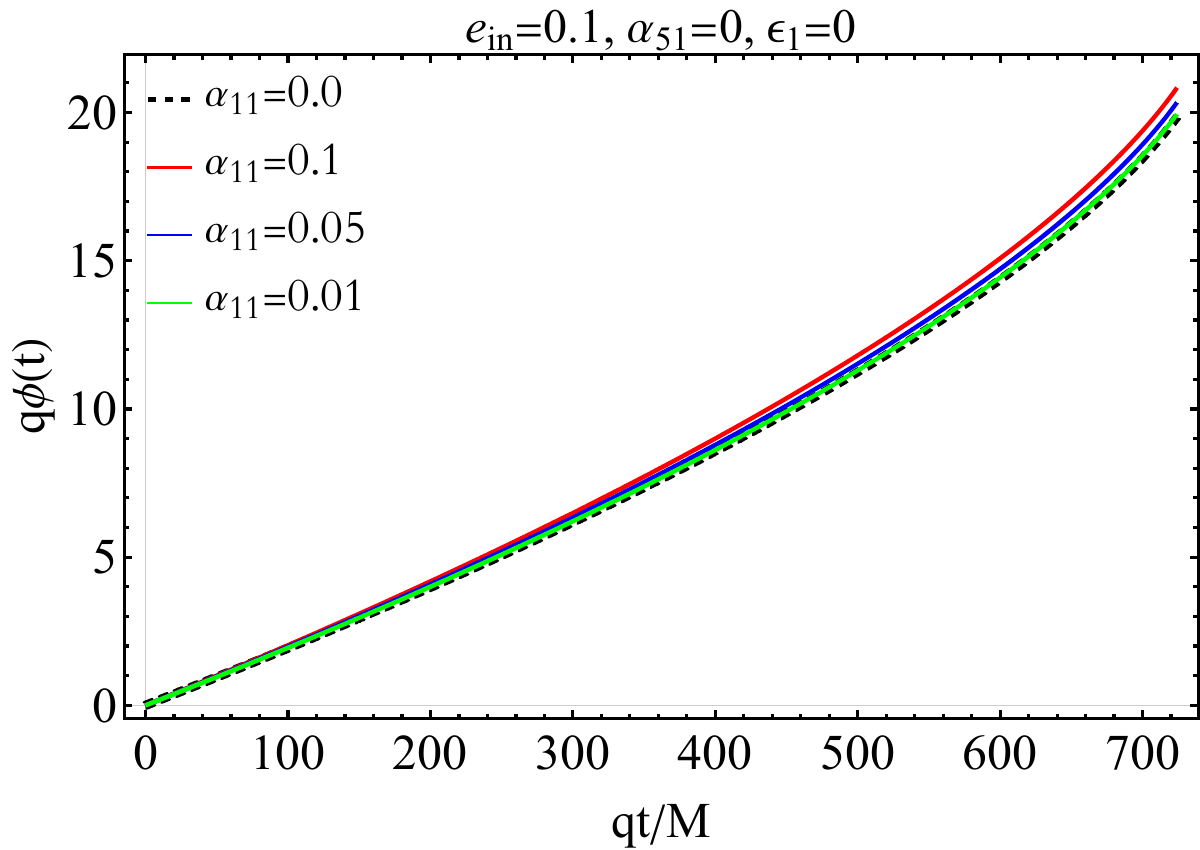}
% \caption{Wormholes for $\Lambda=0$}
	\endminipage\hfill
	%%%%%%%%%%%%%%%%%%%%%%%%
	\minipage{0.33\textwidth}
	\includegraphics[width=\linewidth]{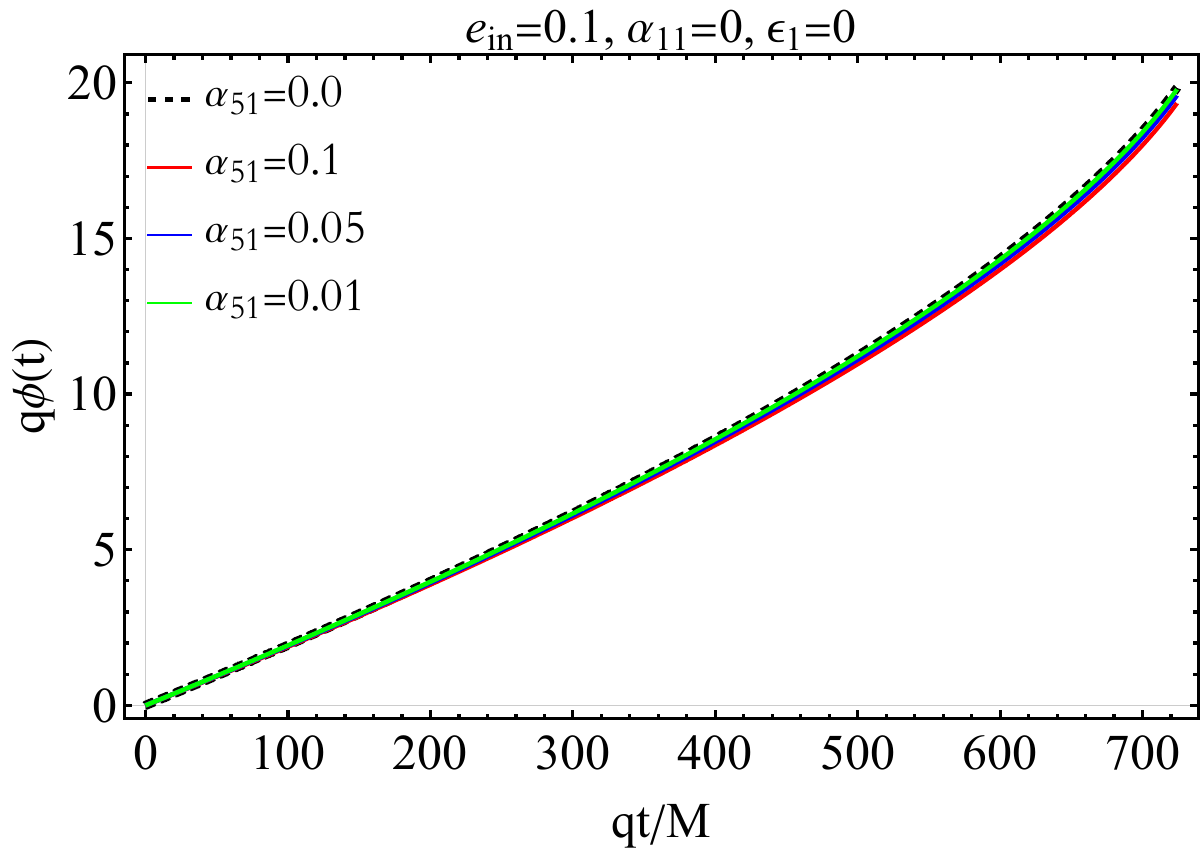}
	\endminipage\hfill
  \minipage{0.33\textwidth}
         \includegraphics[width=\linewidth]{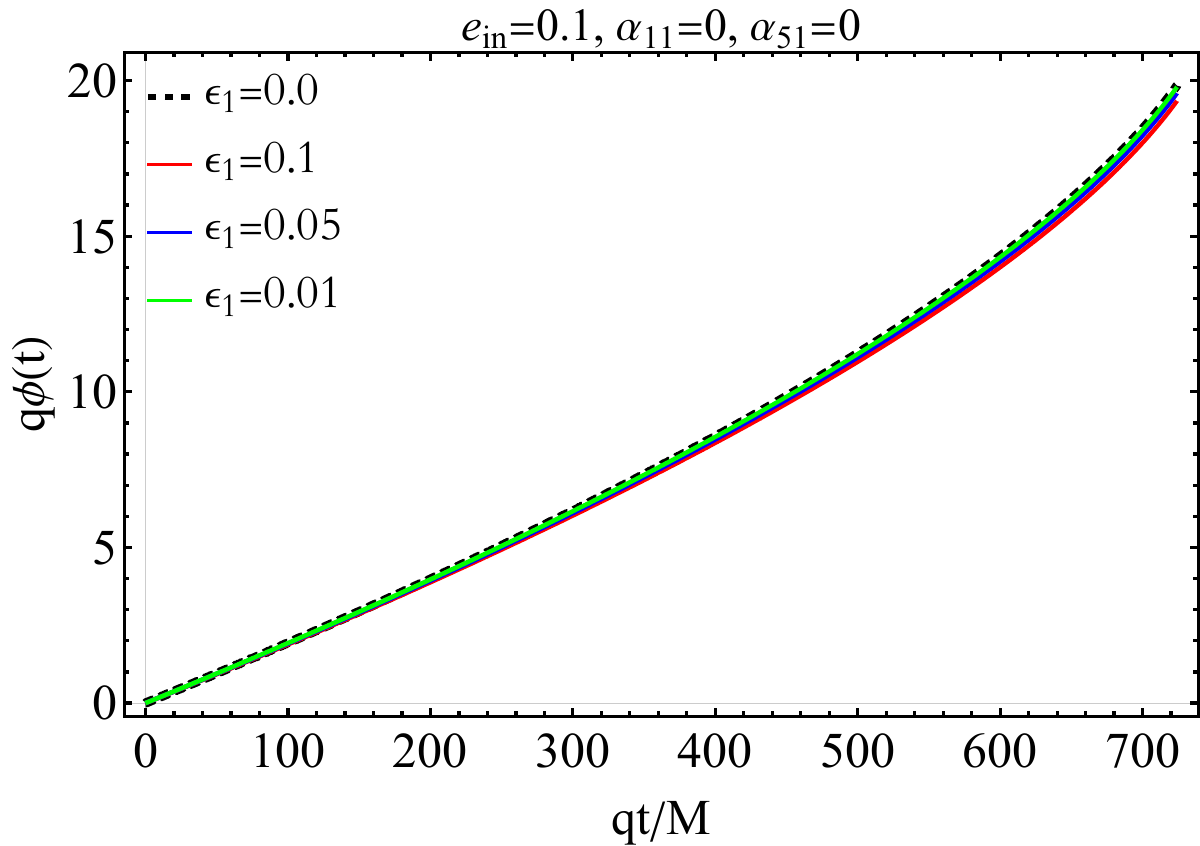}
	\endminipage\hfill
 \minipage{0.33\textwidth}
	\includegraphics[width=\linewidth]{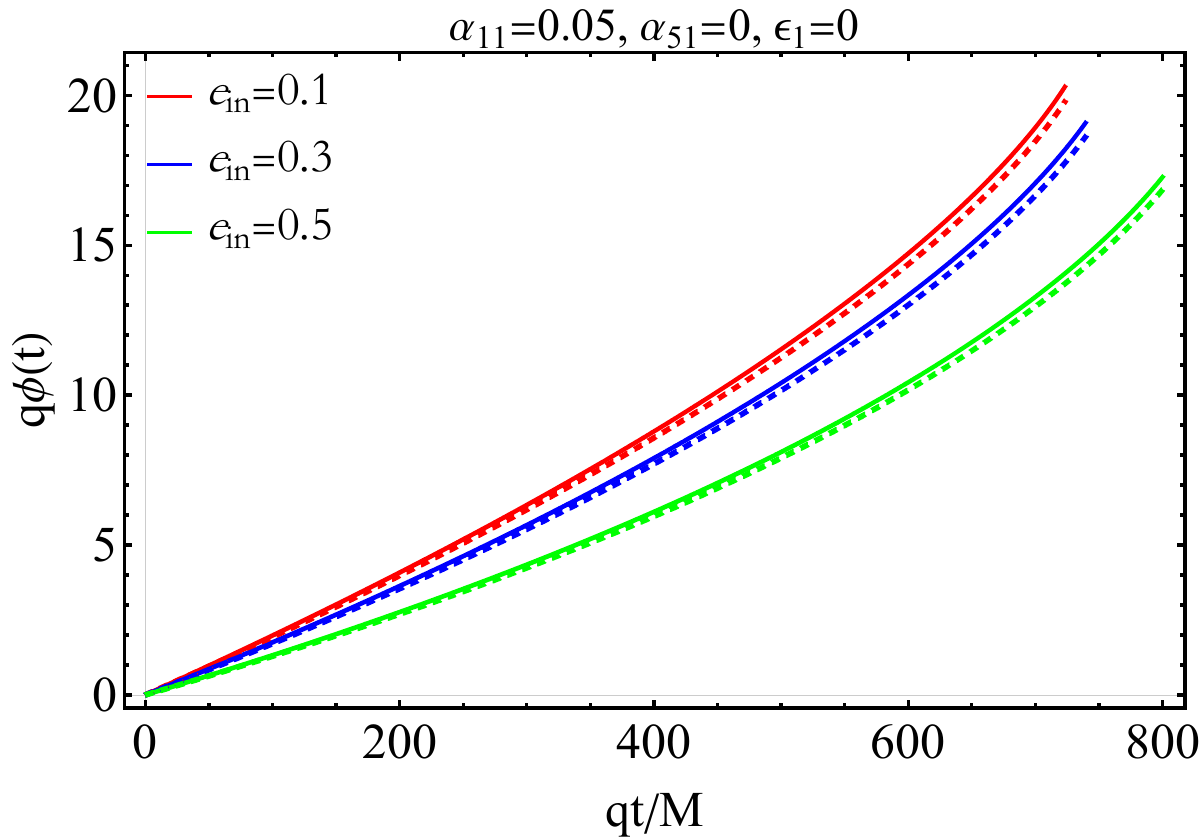}
	\endminipage\hfill
  \minipage{0.33\textwidth}
         \includegraphics[width=\linewidth]{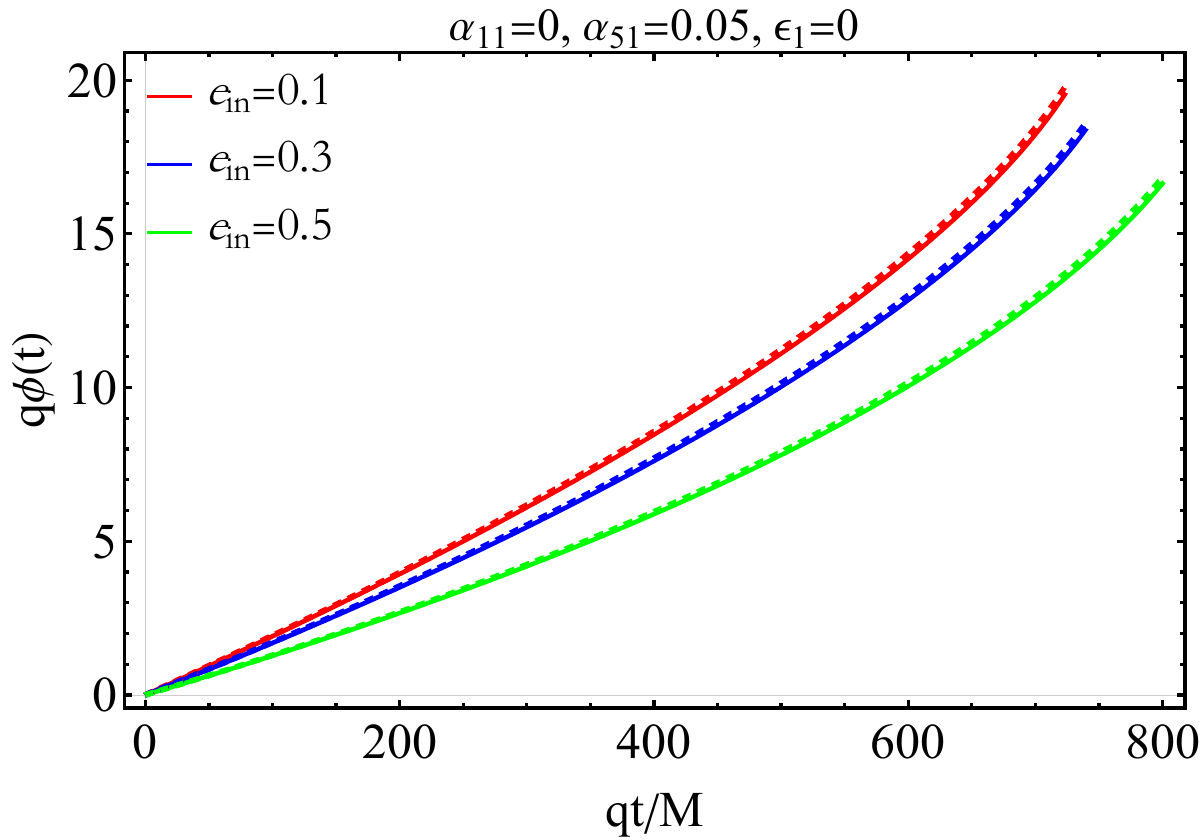}
	 \endminipage\hfill
  \centering
	\minipage{0.33\textwidth}
	\includegraphics[width=\linewidth]{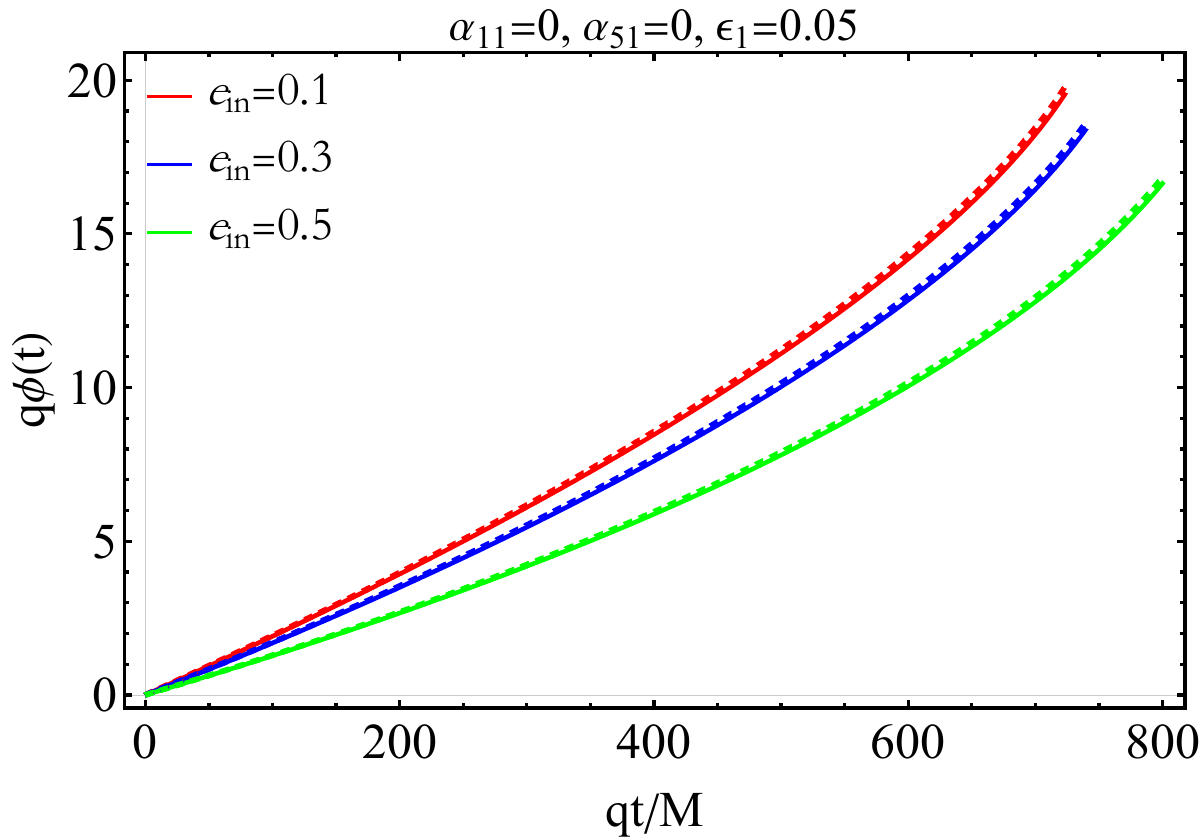}
	\endminipage
	\caption{In the upper panel, we show the time evolution of orbital phase $\phi$ for a fixed initial eccentricity $e_{\textup{in}}=0.1$ and different values of ($\alpha_{11}, \alpha_{51}, \epsilon_{1}$). The dotted curve represents the result without deformation. We consider the starting point of evolution of the inspiralling object $p_{\textup{in}}=14$ and truncate its trajectory at the lowest stable orbit. The lower panel shows the same plots with distinct initial eccentricities and fixed deformation parameters. 
 }\label{fig_enr_flux_same_e}
\end{figure}
Here, functions ($p, e$) depend on time, whose solutions are obtained from Eq. (\ref{dpdtn}). We further solve Eq. (\ref{phase}) to obtain the orbital phase, which will ultimately capture the effects of deviation parameters. As we are focusing on quadrupolar radiation only, we can consider that the GW phase will be twice the orbital phase, $\Phi_{\textup{GW}}(t)=2\phi(t)$. In Fig. (\ref{fig_enr_flux_same_e}), we examine the effect of deformations on orbital phase $\phi(t)$. We find that the larger the value of $\alpha_{11}$, the larger the phase is. However, it goes the other way round for ($\alpha_{51}, \epsilon_{1}$). This nature completely depends on the relative sign of the deviation parameters. We further notice that small initial eccentricities give rise to the larger values of phase shift. We next compute the waveform and estimate the detectability of deformation parameters with LISA observations.
\section{Detectability: Waveform and Mismatch}\label{detect}
Now we turn our discussion to the detectability of deformation parameters with GWs from EMRIs. The presence of deviation parameters will influence the GW signals. This effect will be observable in the gravitational waveforms, altering their characteristics in a detectable manner. 

\subsection{Waveform}We start with generating waveforms for distinct values of deformations appearing as leading order corrections in the derived results.  In the transverse traceless (TT) gauge, the gravitational field, represented by the metric perturbation $h_{ij}^{TT}$, is expressed straightforwardly as \cite{Gopakumar:2001dy, Yunes:2009yz, Dai:2023cft, Babichev:2024hjf}
\begin{align}
h_{ij}^{TT} = \frac{2}{R}\Ddot{I}_{ij}^{TT},
\end{align}
where $R$ is the luminosity distance between detector and source, and $I_{ij}$ is the quadrupole mass moment for the system under consideration. The $h_{ij}^{TT}$ provides us with two tensor modes (polarization modes) ($h_{+}, h_{\times}$) given in terms of two unit vectors ($p, q$) in the transverse subspace of the propagation direction. These vectors allow for a clear separation of the polarization components of the GW signal. With this, the polarization modes can be computed by using \cite{Moore:2016qxz, Gopakumar:2001dy, Yunes:2009yz},
\begin{align}
h_{+} = \frac{1}{2}(p_{i}p_{j}-q_{i}q_{j})h^{TT}_{ij} \hspace{3mm} ; \hspace{3mm} h_{\times} = \frac{1}{2}(p_{i}q_{j}-p_{j}q_{i})h^{TT}_{ij}\,.
\end{align}
Following \cite{Moore:2016qxz, Gopakumar:2001dy}, we write 
\begin{equation}
\begin{aligned}\label{rfg}
h_{+} =& -\frac{2}{R}\Big(\Big(\frac{1}{r}+r^{2}\dot{\phi}^{2}-\dot{r}^{2}\Big)\cos{2\phi}+2r\dot{r}\dot{\phi}\sin2\phi\Big) \\
 h_{\times} =& -\frac{2}{R}\Big(\Big(\frac{1}{r}+r^{2}\dot{\phi}^{2}-\dot{r}^{2}\Big)\sin2\phi-2r\dot{r}\dot{\phi}\cos2\phi\Big),
\end{aligned}
\end{equation}
replacing the orbital velocities using Eqs. (\ref{gdscs3n}), we obtain
\begin{equation}
\begin{aligned}\label{wavform}
h_{+} =& -\frac{2}{pR}\Big(\cos2 \phi (3 e \cos \chi+2)+e (e \cos (2 (\phi -\chi ))+2 \sin\chi \sin2 \phi)\Big) \\
& -\frac{1}{pR}\Big(e^2 \cos (2 (\phi -\chi ))+2 e \cos (2 \phi -\chi )+\cos2 \phi\Big)(2\alpha_{11}-\alpha_{51}-\epsilon_{1})\,, \\
h_{\times} =& -\frac{1}{pR}\Big(e (2 e \sin (2 (\phi -\chi ))+5 \sin (2 \phi -\chi )+\sin (\chi +2 \phi ))+4 \sin 2\phi\Big) \\
& -\frac{1}{pR}\Big(e^2 \sin (2 (\phi -\chi ))+2 e \sin (2 \phi -\chi )+\sin2 \phi\Big)(2\alpha_{11}-\alpha_{51}-\epsilon_{1})\,.
\end{aligned}
\end{equation}
Also, the orbital parameters ($p, e, \phi, \chi$) are functions of time $t$, which can be obtained from Eqs. (\ref{phase}, \ref{dpdtn}, \ref{freq}). Since the expressions in (\ref{wavform}) are already linear in deviations, the time evolution of orbital parameters comes only from the GR part while evaluating as we are working in linear order in these deviation parameters. Note that the expressions are written in dimensionless units, as mentioned earlier. However, we dimensionalize quantities in physical units while performing numerical computations. 

We used Eq. (\ref{rfg}) to generate the waveform signals. We then perform numerical computations to determine the time-domain GW waveforms for different values of deformation parameters and eccentricities and compare the results with the case when deformations are switched off. We present the waveform signals during the initial stages of the inspiral, i.e., for a few hours in the beginning when the inspiral starts and for a few hours toward the end of the inspiral, as depicted in \textcolor{black}{Fig. (\ref{waveform1}) and Fig. (\ref{waveform}), including Fig. (\ref{waveform2}) in appendix (\ref{apendwave}). Fig. (\ref{waveform}) shows the waveforms when a single deviation parameter is switched on and the rest are set to zero, while Figs. (\ref{waveform1}) and (\ref{waveform2}) present the waveforms when all deviations are turned on simultaneously.} %We first start with the analytical expression of the waveform in the frequency domain, and next, we show the time domain waveform with mismatch analysis.

\begin{figure}[h!]
\centering

	\minipage{0.49\textwidth}
	\includegraphics[width=\linewidth]{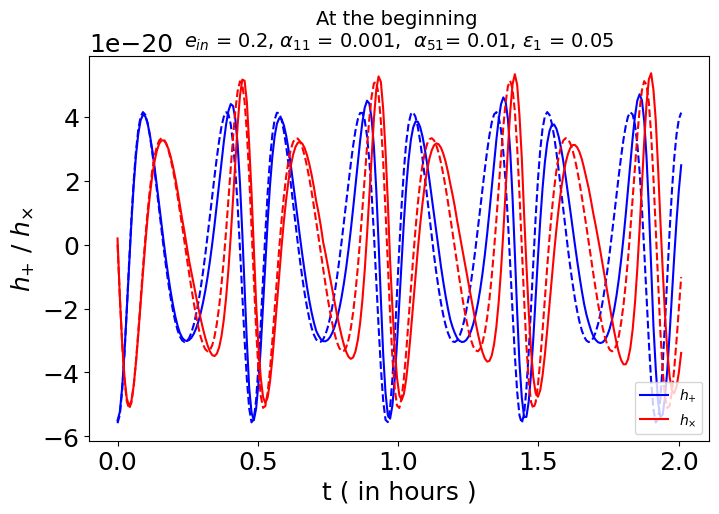}
% \caption{Wormholes for $\Lambda=0$}
	\endminipage\hfill
	%%%%%%%%%%%%%%%%%%%%%%%%
	\minipage{0.49\textwidth}
	\includegraphics[width=\linewidth]{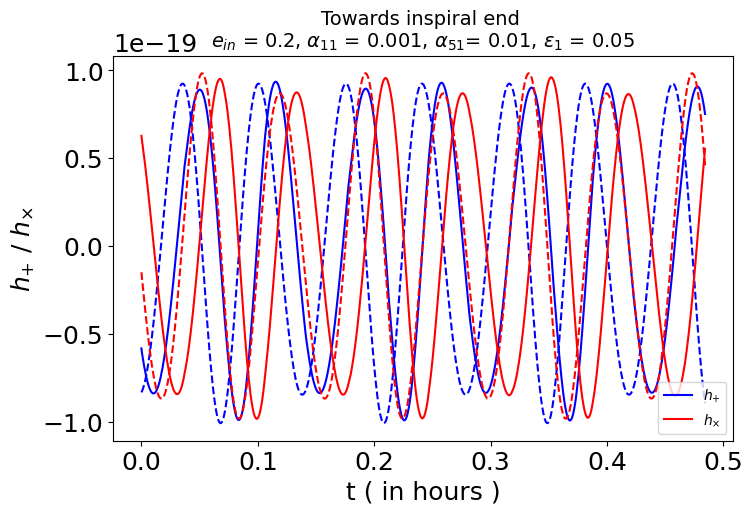}
	\endminipage\hfill
  \minipage{0.49\textwidth}
         \includegraphics[width=\linewidth]{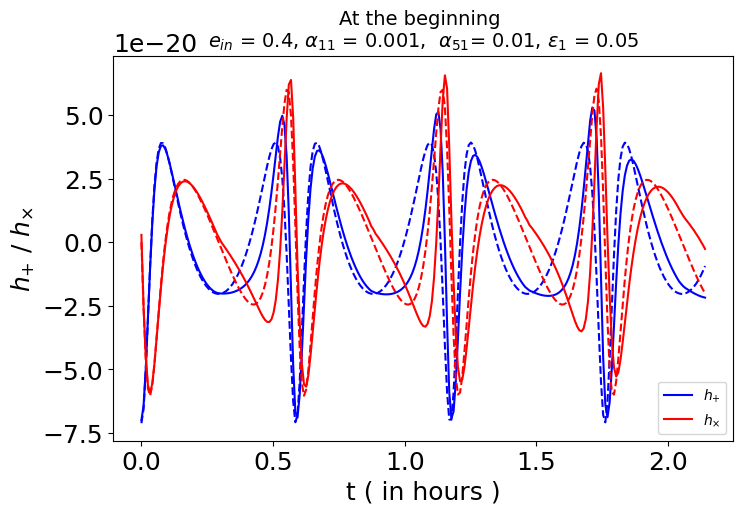}
	\endminipage\hfill
  \minipage{0.49\textwidth}
	\includegraphics[width=\linewidth]{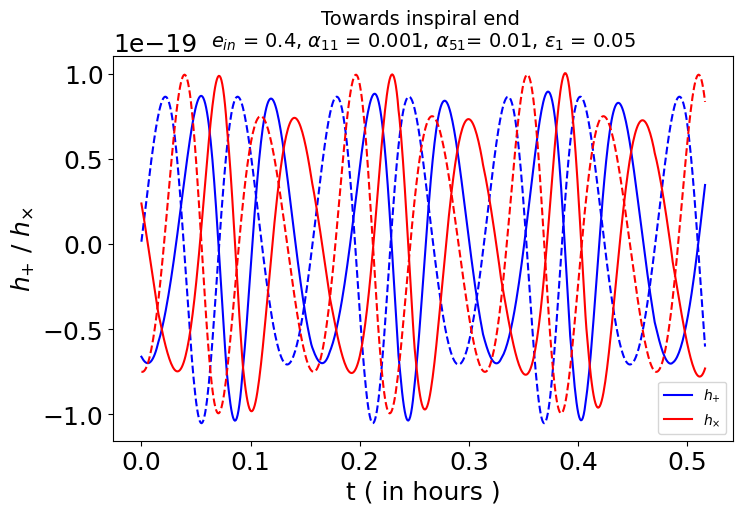}
	\endminipage
 
	\caption{The waveforms represent the GW signal with $\mu=10 M_\odot$ inspiralling the primary $M=10^{6}M_\odot$. The inspiral starts at $p=14$ and luminosity distance $R=10 \textup{Mpc}$. The non-zero deviations have been chosen for all three deviation parameters at $e_{\textup{in}}=0.2$ and $e_{\textup{in}}=0.4$. The left column shows the effects of the deviation parameters on the waveforms ($h_{+}, h_{\times}$) for the initial hours. The right column describes the same for the last few minutes towards the end of the inspiral. The dotted curves depict the GR part of the corresponding signal.} \label{waveform1}

\end{figure}
%%%%%%%%%%%%%%%%%%%%%%%%%%%%%%%%%%%%%%%%%%%%%%%%%%%%%%%%%%%%%%%%%%%%%%%%%%%%%%%%%
\begin{figure}[h!]
\centering
	\minipage{0.49\textwidth}
	\includegraphics[width=\linewidth]{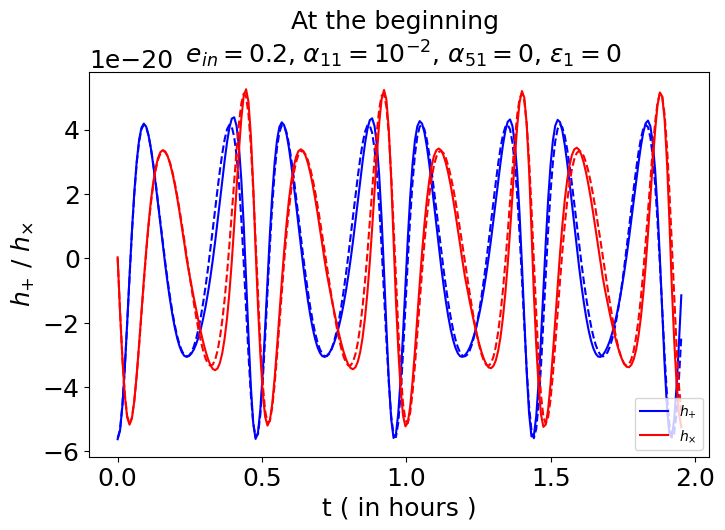}
% \caption{Wormholes for $\Lambda=0$}
	\endminipage\hfill
	%%%%%%%%%%%%%%%%%%%%%%%%
	\minipage{0.49\textwidth}
	\includegraphics[width=\linewidth]{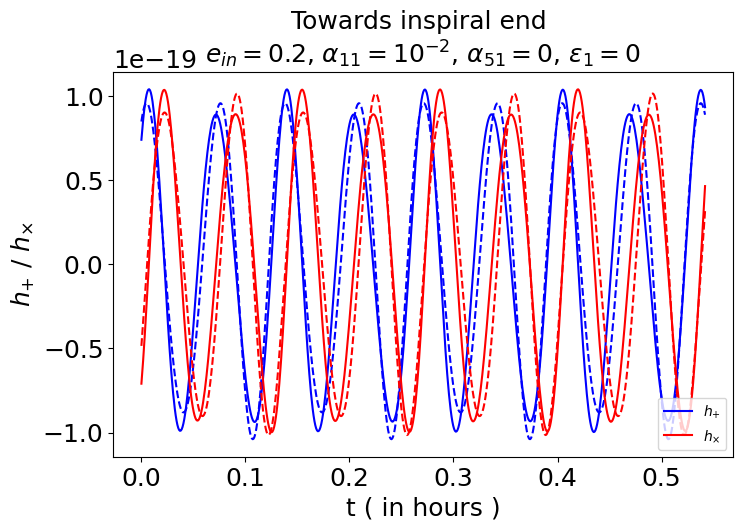}
	\endminipage\hfill
  \minipage{0.49\textwidth}
         \includegraphics[width=\linewidth]{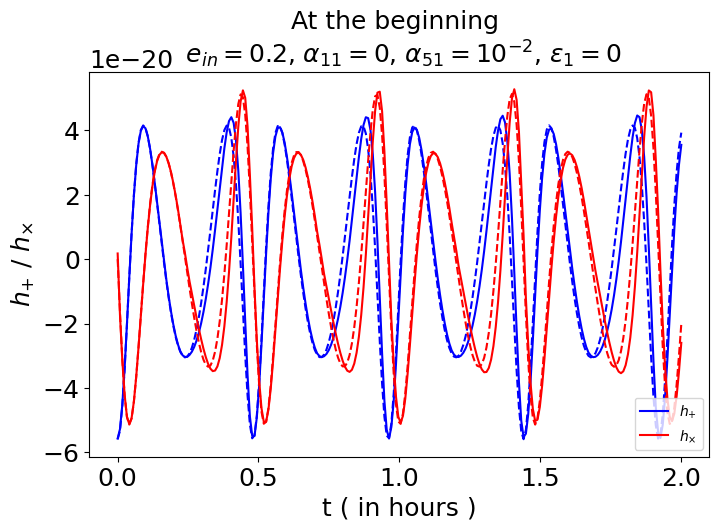}
	\endminipage\hfill
  \minipage{0.49\textwidth}
	\includegraphics[width=\linewidth]{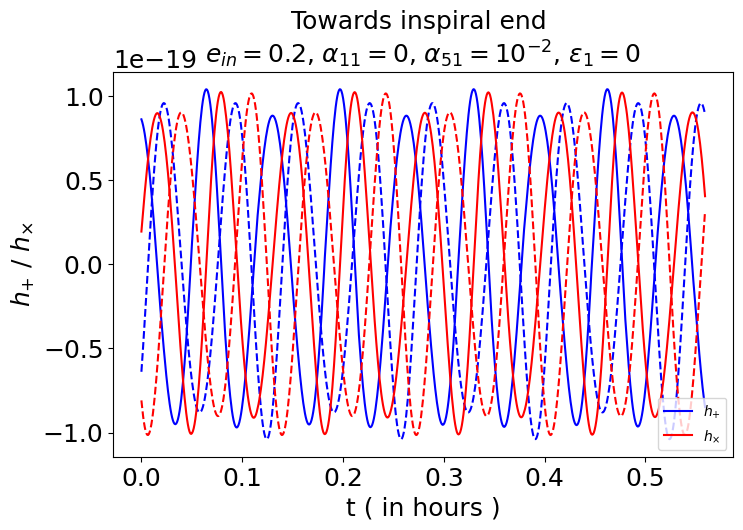}
	\endminipage\hfill
  \minipage{0.49\textwidth}
         \includegraphics[width=\linewidth]{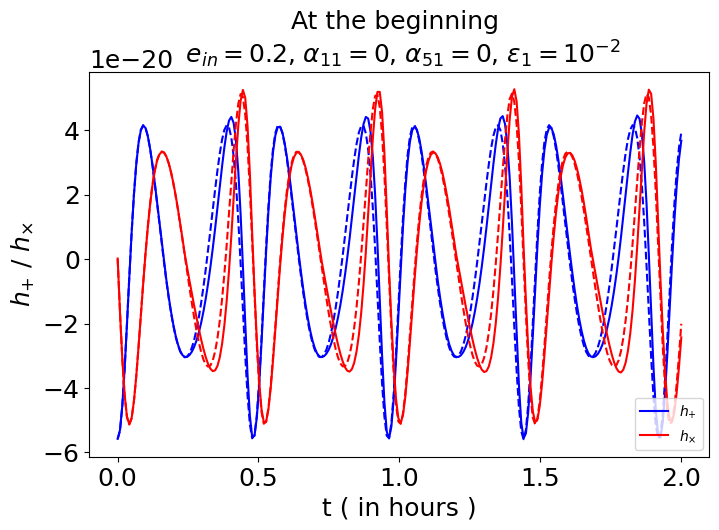}
	 \endminipage\hfill
  \centering
	\minipage{0.49\textwidth}
	\includegraphics[width=\linewidth]{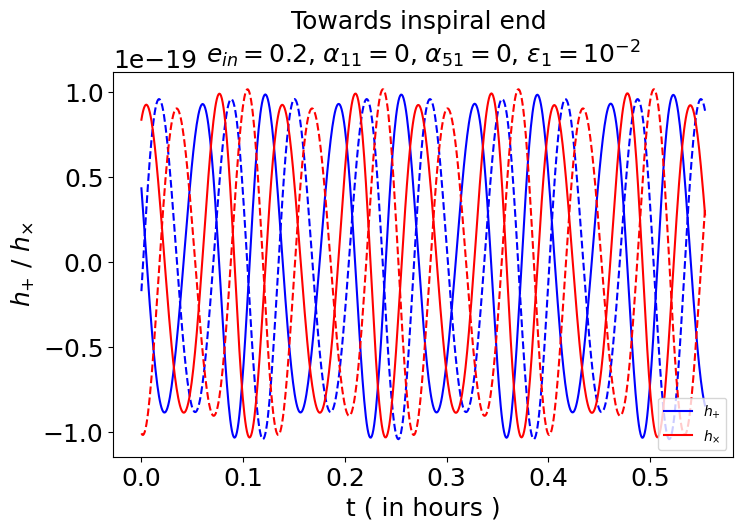}
	\endminipage
% \centering
% 	\minipage{0.33\textwidth}
% 	\includegraphics[width=\linewidth]{Kent_Yagi_Eccentric_PN/alpha11 begin.png}
% % \caption{Wormholes for $\Lambda=0$}
% 	\endminipage\hfill
% 	%%%%%%%%%%%%%%%%%%%%%%%%
% 	\minipage{0.33\textwidth}
% 	\includegraphics[width=\linewidth]{Kent_Yagi_Eccentric_PN/alpha51 begin.png}
% 	\endminipage\hfill
%   \minipage{0.33\textwidth}
%          \includegraphics[width=\linewidth]{Kent_Yagi_Eccentric_PN/epsilon1 begin.png}
% 	\endminipage\hfill

%   \minipage{0.33\textwidth}
% 	\includegraphics[width=\linewidth]{Kent_Yagi_Eccentric_PN/alpha11 end.png}
% 	\endminipage\hfill
%   \minipage{0.33\textwidth}
%          \includegraphics[width=\linewidth]{Kent_Yagi_Eccentric_PN/alpha51 end.png}
% 	 \endminipage\hfill
%   \centering
% 	\minipage{0.33\textwidth}
% 	\includegraphics[width=\linewidth]{Kent_Yagi_Eccentric_PN/epsilon1 end.png}
% 	\endminipage

	\caption{The waveforms represent the GW signal with $\mu=10 M_\odot$ inspiralling the primary $M=10^{6}M_\odot$. The inspiral starts at $p=14$ and luminosity distance $R=10 \textup{Mpc}$. The deviations have been chosen as $10^{-2}$ and $e_{\textup{in}}=0.2$. The left column shows the effects of the deviation parameters on the waveforms ($h_{+}, h_{\times}$) for the initial hours. The right column describes the same for the last few minutes towards the end of the inspiral. The dotted curves depict the GR part of the corresponding signal.} \label{waveform}
% 	%%%%%%%%%%%%%%%%%%%%%%%%
% 	\centering
% 	\minipage{0.5\textwidth}
% 	\includegraphics[width=\linewidth]{Kent_Yagi_Eccentric_PN/alpha11 begin.png}
% % \caption{Wormholes for $\Lambda=0$}
% 	\endminipage\hfill
% 	%%%%%%%%%%%%%%%%%%%%%%%%
% 	\minipage{0.5\textwidth}
% 	\includegraphics[width=\linewidth]{}
% 	\endminipage\hfill
%   \minipage{0.5\textwidth}
%          \includegraphics[width=\linewidth]{Kent_Yagi_Eccentric_PN/alpha51 begin.png}
% 	\endminipage\hfill
%  \minipage{0.5\textwidth}
% 	\includegraphics[width=\linewidth]{}
% 	\endminipage\hfill
%   \minipage{0.5\textwidth}
%          \includegraphics[width=\linewidth]{Kent_Yagi_Eccentric_PN/epsilon1 begin.png}
% 	 \endminipage\hfill
%   \centering
% 	\minipage{0.5\textwidth}
% 	\includegraphics[width=\linewidth]{}
% 	\endminipage
% 	\caption{}\label{}
\end{figure}
%%%%%%%%%%%%%%%%%%%%%%%%%%%%%%%%%%%%%%%%%%%%%%%%%%%%%%%%%%%%%%%%%%%%%%%%%%%%%%
%%%%%%%%%%%%%%%%%%%%%%%%%%%%%%%%%%%%%%%%%%%%%%%%%%%%%%%%%%%%%%%%%%%%%%%%%%%%%%%

Furthermore, one can analyze the counterpart of the time-domain chirp signal in the frequency domain, a method that is also commonly applied in GW data analysis. One can derive an analytic expression for the GW phase in the frequency domain within the stationary phase approximation ($\Psi_{\textup{SPA}}$) and, correspondingly, the waveforms. \textcolor{black}{Since many studies focus on non-GR deviation primarily in the phase as detectors tend to exhibit greater sensitivity to the deviation in phase, let us compute such a quantity now. We will be ignoring the effect of deviation parameters in the amplitude of strain.} As we are considering quadrupolar radiation, the GW frequency is $f\equiv f_{GW}=\Omega/\pi$; where $\Omega=d\phi/dt$ is given by Eq. (\ref{phase}). Hence, we can write down
\begin{align}\label{freqD1}
f = \frac{(1-e)^{3/2}}{4M\pi p^{3/2}}(4+2\alpha_{11}-\alpha_{51}-\epsilon_{1}).
\end{align}
One can also invert the $dp/dt$ in Eq. (\ref{dpdtn}) and obtain
\begin{align}
p(\Omega) = \frac{(1-e^{2})}{6(M\Omega)^{2/3}}(6+2\alpha_{11}-\alpha_{51}-\epsilon_{1}).
\end{align}
Now derivating Eq. (\ref{freqD1}) with respect to $t$ and using Eq. (\ref{dpdtn}), we arrive at the following expression:
\begin{align}
\frac{df}{dt} = \frac{96}{5} \frac{\widetilde{\mathcal{M}}^{5/3} f^{11/3}\pi^{8/3}}{(1-e^{2})^{7/2}}\Big(1+\frac{1}{3}(2\alpha_{11}-\alpha_{51}-\epsilon_{1})\Big).
\end{align}
We can further integrate the above equation for a fixed eccentricity, $e_{0}$, and obtain
\begin{align}\label{time1}
t = t_{c}-\frac{5(\pi f)^{-8/3}}{W(e_{0})}\Big(1+\frac{1}{3}(2\alpha_{11}-\alpha_{51}-\epsilon_{1})\Big),
\end{align}
where $t_{c}$ is the coalescence time; here, in particular, it can be referred as the the region where the secondary object truncates its trajectory. Further, an analytical expression for the phase of the waveform using $\Phi_{GW} = \int^{t}2\pi f(t')dt'$ \cite{Cutler:1994ys} can be derived,
\begin{align}
\Phi_{GW} (t)= -2W(e_{0})^{-3/8}\Big(\frac{\widetilde{\mathcal{M}}^{-1}}{5}(t_{c}-t)\Big)^{5/8} \Big(1-\frac{1}{8}(2\alpha_{11}-\alpha_{51}-\epsilon_{1})\Big),
\end{align}
where $W(e_{0})\equiv\frac{1}{(1-e_{0}^{2})^{7/2}}(1+\frac{292}{96}e_{0}^{2}+\frac{37}{96}e_{0}^{4})$ and $\widetilde{\mathcal{M}}\equiv\mu^{3/5}M^{2/5}$. $e_{0}$ represents the fixed initial eccentricity. It is clear that if we switch off the deviations and set $e_{0}=0 \Longrightarrow W(e_{0})=1$, the analysis gets reduced to circular from eccentric, which is in agreement with \cite{Cutler:1994ys}. Further, we obtain the following relations:
\begin{align}\label{time2}
\Phi_{GW}(f) = \Phi_{c}-\frac{(\pi f \widetilde{\mathcal{M}})^{-5/3}}{16W(e_{0})}\Big(1+\frac{1}{12}(2\alpha_{11}-\alpha_{51}-\epsilon_{1})\Big).
\end{align}
The $\Phi_{c}$ is the value of the phase at the coalescence time, also termed the coalescence phase. In other words, as the coalescence approaches, the GW frequency gets larger; thus, $f\longrightarrow\infty \Longrightarrow (t\longrightarrow t_{c}, \Phi_{GW}\longrightarrow\Phi_{c})$ \cite{Cutler:1994ys, Maggiore:2007ulw}. However, in real scenario, the inspiral will eventually end at a finite orbital frequency.

Further, in order to derive the analytical expression, one can treat the signal (amplitude) as evolving at a much slower rate than the phase and that phase does not vary rapidly, i.e., the stationary phase approximation ($\Psi_{\textup{SPA}}$). Such conditions are maintained within a small eccentricity approximation. With this, the phase is given by $\Psi_{\textup{SPA}} = 2\pi ft-\Phi_{GW}-\frac{\pi}{4}$. Readers are suggested to refer \cite{Damour:2000gg, Droz:1999qx} for further details on SPA. Consequently, in general, the waveform will take the following form,
\begin{align}
\tilde{h}(f) = \mathcal{B}e^{i\Psi_{\textup{SPA}}},
\end{align}
where, $\tilde{h}(f)$ is the Fourier transformation of time domain signal, $h(t)=F_{+}h_{+}+F_{\times}h_{\times}$. Here, ($F_{+}, F_{\times}$) are the antenna pattern functions, which can be found in \cite{Huerta:2011kt, Moore:2016qxz}. However, it is not of concern at present as it sits in the amplitude ($\mathcal{B}$) part, following \cite{Moore:2016qxz}. As a result, we obtain \cite{Moore:2016qxz, Cutler:1994ys}
%Following \cite{Cutler:1994ys, Maggiore:2007ulw, Moore:2016qxz, Isoyama:2020lls}, in general for a given time domain waveform $h(t)=A(t)e^{-i\phi(t)}$, one can treat the signal as evolving at a much slower rate than the phase, i.e., stationary phase approximation ($\Psi_{\textup{SPA}}$): $d(lnA)/dt <<d\Phi_{GW}(t)/$. With this, the phase is $\Psi_{\textup{SPA}} = 2\pi ft-\Phi_{GW}-\frac{\pi}{4}$, which results in the following expression:
\begin{align}
\mathcal{B} =& -4M \sqrt{\frac{5\pi}{96}}\Big(\frac{M}{R}\Big)q^{1/2}(\pi M f)^{-7/6}(F_{+}^{2}+F_{\times}^{2})^{1/2}\,, \\
\Psi_{\textup{SPA}} =& 2\pi ft_{c}-\Phi_{c}-\frac{\pi}{4}+\frac{3}{128}(\pi \widetilde{\mathcal{M}}f)^{-5/3}\Big(1+\frac{157 e_{0}^2}{24}+\frac{605 e_{0}^4}{32}\Big)\Big(1-\frac{1}{3}(2\alpha_{11}-\alpha_{51}-\epsilon_{1})\Big),
\end{align}
where, for deriving $\Psi_{\textup{SPA}}$, we have used Eqs. (\ref{time1}, \ref{time2}) with the fixed small eccentricity ($e_{0}$) expansion. $q$ is the mass-ratio of the system as mentioned earlier. Thus, the $\Psi_{\textup{SPA}}$ contributes to the waveform in the frequency domain with the effects of deviations. Note that the derived results are consistent with \cite{Cutler:1994ys, Maggiore:2007ulw, Moore:2016qxz, Isoyama:2020lls} when deformations are set to zero.

\subsection{Mismatch}
Using time domain EMRI waveforms, we further estimate the mismatch ($\mathcal{M}$) to provide an order of magnitude on the deviation parameters for observing them in LISA. To start with, let us proceed with computing the overlap between two GW signals. When considering these signals $h_{1}(t)$ and $h_{2}(t)$, the overlap is defined as \cite{PhysRevD.78.124020, Babak:2006uv, Rahman:2022fay}
\begin{align}
O(h_{1}, h_{2}) = \frac{(h_{1}\vert h_{2})}{\sqrt{(h_{1}\vert h_{1})(h_{2}\vert h_{2})}}\,.
\end{align}
Their inner product is expressed as
\begin{align}
(h_{1}\vert h_{2}) = 4Re \int_{0}^{\infty}\frac{\Tilde{h}_{1}(f)\Tilde{h}^{*}_{2}(f)}{S_{n}f}df\,, 
\end{align}
where $\Tilde{h}(f)$ is calculated in frequency domain, i.e., Fourier transform of time domain waveform $h(t)$. The $S_{n}(f)$ is the power spectral density whose explicit expression can be found in \cite{PhysRevD.69.082005, PhysRevD.78.124020, Huerta:2011kt}. Once we have the overlap $O$, we define the Mismatch as 
\begin{align}
\mathcal{M} = 1-O(h_{1}, h_{2}).
\end{align}
If two signals are identical, then $O=1$, as a result, the mismatch will be zero. The $\mathcal{M}$ value lies in the range zero to one.
\begin{figure}[h!]
\centering
\includegraphics[width=3.17in, height=2.2in]{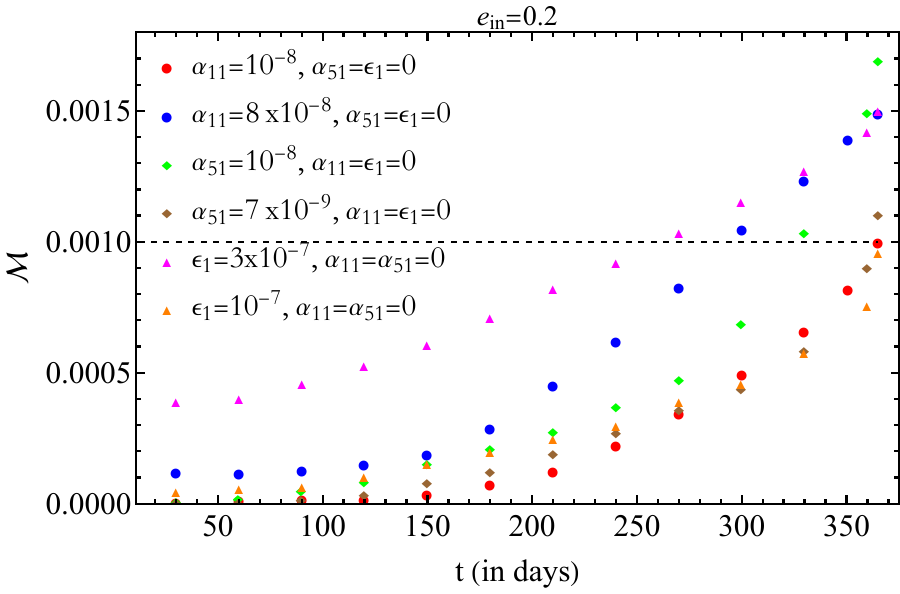}
\includegraphics[width=3.17in, height=2.08in]{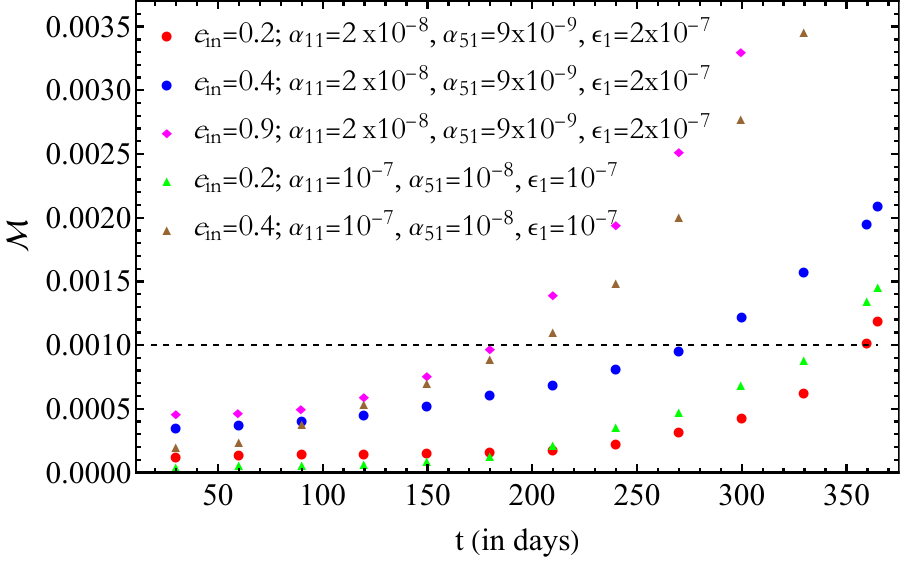}
\caption{Waveform mismatch for the EMRI system under consideration with a primary black hole ($10^{6}M_{\odot}$) and a secondary object ($10M_{\odot}$), showing an estimate on the order of magnitude crossing the detection threshold with the observation period of one year.} \label{mismatch}
\end{figure}

In Fig.~(\ref{mismatch}), we assess the feasibility of probing the deformations through LISA observations, offering an approximate estimation of their potential magnitude. We adhere to a criterion that indicates that gravitational waveforms $h_{1}$ and $h_{2}$ are considered indistinguishable in GW observations when $\mathcal{M}(h_{1}, h_{2})\leq \mathcal{M}_{\textup{threshold}} \approx 1/2\rho^{2}$, where $\mathcal{M}_{\textup{threshold}}$ represents the detection threshold and $\rho$ is the signal-to-noise ratio (SNR). We take $\mathcal{M}_{\textup{threshold}}\approx0.001$ \cite{PhysRevD.78.124020, Rahman:2022fay, Huerta:2011kt} and provide an order of magnitude analysis to constrain the deviation parameters to observe them through LISA. We consider the last stage of the inspiral and observe for a year towards the end of the inspiral. In the left panel of Fig.~(\ref{mismatch}), we keep only one deviation parameter non-zero at a time for a fixed initial eccentricity. In contrast, the right panel of Fig.~(\ref{mismatch}), depicts the scenario in which all deviations are turned on simultaneously. In particular, with $e_{\textup{in}}=0.2\,,$ the values of the deviation parameters for which the estimated mismatch exceeds the detection threshold are $\alpha_{11} \sim 2\times 10^{-8}$, $\alpha_{51} \sim 9\times 10^{-9}$ and $\epsilon_{1} \sim 2\times 10^{-7}$. Moreover, it is clear from the left panel of Fig. (\ref{mismatch}) that the order of magnitude of the deviations remains the same even if only one deviation is taken to be non-zero and the rest are set to zero. 

It is also apparent from both mismatch plots in Fig.~(\ref{mismatch}), that higher initial eccentricities or values of the deviation parameters lead to increased mismatch values, suggesting enhanced detectability through LISA observations. Interestingly, from the plots in the right panel of  Fig.~(\ref{mismatch}), it can be further noted that for a fixed $e_{\textup{in}}$, a higher value of the deviation parameter increases the mismatch at a later stage of the observation. For instance, at $e_{\textup{in}}=0.2, \hspace{0.22mm}\textup{deviations} \hspace{0.17cm} (\alpha_{11} = 10^{-7}, \alpha_{51} = 10^{-8}, \epsilon_{1}=10^{-7})$ yield larger mismatch than deviations $(\alpha_{11} = 2\times 10^{-8}, \alpha_{51} = 9\times 10^{-9}, \epsilon_{1}=2\times 10^{-7})$, implying that higher deviation values lead to increased mismatch as days progress, despite of the fact that initially there are smaller mismatch compared to lower deviation values. Such an aspect completely depends on the relative signs of the deviation parameters. However,  in scenarios where only one deviation parameter is activated while the others are set at zero, as shown in the left panel of Fig.~(\ref{mismatch}), the mismatch behaviour is always increasing in time, as expected.

Further, a thorough parameter estimation utilizing a Fisher's/Bayesian framework can be performed to assess how well LISA can measure the parameters of an EMRI source. These methodologies will facilitate the identification of parameter degeneracies more precisely and enable stringent constraints on non-GR deviations. Briefly, in the Bayesian framework, the posterior distribution of the model parameters captures our understanding of the parameters \cite{Thrane:2018qnx, Cornish:2014kda}. This will undoubtedly help uncover stronger constraints on parameters, including non-GR deviations addressed in the current article. We plan to do a focused analysis of binary parameters as well as these deviation parameters using Fisher's/Bayesian methods in a separate future study, focusing solely on the order-of-magnitude analysis in this current paper.

%The first step in determining this posterior distribution is to use the prior information through the prior distribution. Now, once we know each parameter's acceptable ranges and potential deviations, we can calculate the likelihood by evaluating the model and assessing how well it fits the data. This process allows us to extract quantitative estimates of each parameter from the posterior distribution, along with their uncertainties and other statistical metrics. Detailed visualizations can then illustrate the intricate structures and relationships among parameters \cite{Thrane:2018qnx, Cornish:2014kda}. 
%%%%%%%%%%%%%%%%%%%%%%%%%%%%%%%%%%%%%%%%%%%%%%%%%%%%%%%%%%%%%%%%%%%%%%%%%%%%%%
\section{Discussion}\label{dscn}
EMRI systems, distinguished by their unique characteristics, have attracted significant attention within the scientific community. The study of such binaries encompasses an exploration of their detectability as sources and their potential for testing gravitational theories, with a particular emphasis on future space-based detectors. Developing highly accurate waveform templates that intricately consider the effects of radiation backreaction within the dynamics of EMRIs is of utmost importance. The comprehensive strategy is crucial for enhancing our ability to detect GW signals originating from such binaries. With this motivation, we attempt to probe the deformations as leading order effects to relativistic PN results. Probing these deformations indicates valuable insight into theories beyond GR, as such parameters can be mapped to many other theories in modified gravity. Let us briefly touch upon the summary of the investigations.

We consider the primary supermassive black hole ($M=10^{6}M_\odot$) that reflects deviation parameters ($\alpha_{11}, \alpha_{51}, \epsilon_{1}$) in the ultimate results as a leading order PN effect. We do not impose ppN constraints here, as Birkhoff's theorem is not guaranteed to hold in general for non-GR theories \cite{Yagi:2023eap}. With this motivation, we first determine the geodesic motion of the inspiralling object exhibiting equatorial eccentric dynamics and find the separatrix region where the secondary truncates its trajectory. We should recall that we take the mass-ratio $q=10^{-5}$ and start the inspiral of the secondary object at $p=14$. We determine the expressions for rate change of orbital energy and angular momentum that relate to GW fluxes within the adiabatic approximation. These fluxes enable us to obtain the orbital motion of the inspiralling object and estimate the timescale difference arising from deformations to reach the LSO. One needs to be careful here with the flux computation. As mentioned earlier in the footnote around (\ref{blnceqn}), we only consider the leading order GR piece for the flux computation, with the non-GR piece being ignored since it is a perturbative correction. To include a non-GR piece in the flux, we would need to rely on a particular theory, handle the fundamental fields, and perform a systematic PN analysis to produce the corresponding results, something that is not in the scope of this paper. We are primarily focused on a systematic analytical approach that encompasses various non-GR results through the mapping to test the possible deviation from GR. In order to have a more stringent bound, one should include non-GR emissions.
\par 
In exploring non-GR EMRI scenarios, one can introduce these non-GR corrections either by modifying the geometry of the primary source or by altering the characteristics of the secondary object or both. For example, as mentioned in \cite{Maselli:2021men}, the secondary could be charged, with the charge being generated by a scalar field, but the primary is taken to be a solution of GR only. In our case, we chose the latter and incorporated these parameters by modifying the geometry of the central object (primary object in the EMRI system), which in turn affects the geodesic equations that govern the motion of the secondary object. We further calculate the impacts of the deformation parameters on the orbital phase with the effects of different initial eccentricity values. Next, from the detectability perspective, we generate GW waveforms with $e_{\textup{in}}=0.2$ and deformations of $10^{-2}$ order. We present the waveform characteristics during both the initial and final few hours and compare them with waveforms generated without deformations. The latter are represented by dotted curves for a clear comparison in Fig. (\ref{waveform}). We also show the behaviour of waveform when all deviations are switched on, given in Figs. (\ref{waveform1}, \ref{waveform2}). Lastly, to constrain the deformation parameters, we use these waveforms to estimate the mismatch for the last one year and determine the possible detectability of deviations with LISA observations. This implies that if LISA collects the data for a year during the last stage of the inspiral, it can detect the deformations to GR within the leading order relativistic PN corrections. We determine, given the detection threshold as $\mathcal{M}_{\textup{threshold}}\approx 10^{-3}$, an order of magnitude bound for deformations as ($\alpha_{11}\sim 2\times 10^{-8}, \alpha_{51}\sim 9\times 10^{-9}, \epsilon_{1}\sim 2\times 10^{-7}$). These small values appear as a bound on the deformations, enabling their detectability with LISA. Thus, the study constrains the parameters that bring deviations to GR with possible LISA observations.

Since the study indicates the deviations emerging at the zero PN as a leading order correction, we next aim to generate the full waveform analysis up to 2PN, where many other deviation parameters will also contribute with their higher-order corrections. We anticipate that the inclusion of higher-order PN corrections will offer valuable insights into such analyses, a topic that we intend to explore further in our forthcoming studies. As we previously reported an analysis when holding ppN constraints \cite{AbhishekChowdhuri:2023gvu}, where we determined that the leading order deformations will appear at 2PN, in the future investigation, we aim to jointly present a general study that includes both considerations with and without ppN constraints. Further, in order to give more detailed insights into deformations mapped to parameters of different theories, we also plan to conduct a Fisher's and Bayesian analysis to more strictly constrain these deviations \cite{Vallisneri:2007ev, LIGOScientific:2021sio}. Black hole perturbation is another way to investigate these scenarios \cite{Li:2022pcy, Pound:2021qin}, which could considerably enhance the precision of PN results and aid in constraining deviation parameters with potential detectability. We hope to present some of these studies in the near future. If LISA could potentially establish compelling limits on deviations or deformations from GR, it presents an intriguing avenue that could be effectively investigated through the lens of EMRIs with deformed black hole spacetimes.

%One of the key research objectives of the LISA mission is to map Kerr spacetime \cite{Glampedakis_2006}. To conduct such a major test of GR, we need to be able to set up the required tools that can measure any deviations from the Kerr metric. This will offer deep insights into the weak and strong gravity regimes of the supermassive black hole (SMBH) that can capture a stellar-mass object which emits gravitational radiation while inspiralling the SMBH. The generated GWs carry information of the spacetime geometry, hence acting as a probe for the GR test. In this direction, EMRIs are the perfect astrophysical objects for testing GR with LISA; thus, the comprehensive mapping of the spacetime metric, where the inspiralling object moves, should be made possible by the detection of GWs from EMRIs \cite{Glampedakis_2005}. Here, we attempt to analyze such a notion, which brings the deviations in Kerr metric and can have possible detectability through low-frequency detectors.
%%%%%%%%%%%%%%%%%%%%%%%%%%%%%%%%%%%%%%%%%%%%%%%%%%%%%%%%%%%%%%%%%%%%%%%%%%%%%%%%%%%%%%%%%%%%%%%%%%%

\section*{Acknowledgements} 
The authors would like to thank Badri Krishnan and Enrico Barausse for valuable insights and also to Mostafizur Rahman and Abhishek Sharma for various useful discussions. The research of S. K. is funded by the National Post-Doctoral Fellowship (N-PDF) from SERB, DST, Government of India (PDF/2023/000369). R. K. S. acknowledges the financial assistance by Sabarmati Bridge Fellowship (MIS/IITGN-SBF/PHY/AB/2023-24/023) from IIT Gandhinagar. A. C. is supported by the Prime Minister Research Fellowship (PMRF-192002-1174) of the Government of India. A.B is supported by the Core Research Grant (CRG/2023/005112) by the Department of Science and Technology Science and Engineering Research Board (India). We also thank the participants of the (virtual) workshop ``Testing Aspects of General Relativity-II", ``Testing Aspects of  General Relativity-III" and ``New insights into particle physics from quantum information and gravitational waves" at Lethbridge University, Canada, funded by McDonald Research Partnership-Building Workshop grant by the McDonald Institute for valuable discussions. A.C. would like to express gratitude to the Observatoire de Paris, LUTH, Meudon for their warm hospitality during much of this work, as well as during the writing of the initial draft of the manuscript. The authors would also like to thank the anonymous referee for useful comments and suggestions.
%%%%%%%%%%%%%%%%%%%%%%%%%%%%%%%%%%%%%%%%%%%%%%%%%%%%%%%%%%%%%%%%%%%%%%%%%%%%%%%%%%%%%%%%%%%%%%%%%%

\appendix

\section{Geodesic equation and constants of motion} \label{apenteu1}
Here, we express the 4-velocity of the inspiralling object that is hovering in the vicinity of the primary deformed Kerr background. Our ultimate analysis considers the equatorial orbits; however, we keep the calculations general in this particular section and set the Carter constant ($\mathcal{Q}$) to zero at later stages. One can employ the Hamilton-Jacobi techniques with the given action and equation \cite{Yagi:2023eap}
\begin{align}\label{ac1}
S = -\frac{1}{2}\mu^{2}\tau-Et+J_{z}\phi+R(r)+\Theta(\theta) \hspace{3mm} ; \hspace{3mm} -\frac{\partial S}{\partial\tau} = \frac{1}{2}g^{\mu\nu}\frac{\partial S}{\partial x^{\mu}}\frac{\partial S}{\partial x^{\nu}}\,,
\end{align}
%The separability of the geodesic motion will be governed by the Hamilton-Jacobi equation \cite{Johannsen:2013szh, Staelens:2023jgr} 
%\begin{align}\label{ac2}
%\frac{1}{2}g^{\mu\nu}\partial_{\mu}S\partial_{\nu}S=-\frac{\partial S}{\partial\tau}\,.
%\end{align}
% Using the metric (\ref{metric}), Eq.(\ref{ac1}) and Eq.(\ref{ac2}), we recover separable equations in radial and angular terms which take the following form,
and the separable equations in radial and angular terms take the following form
\begin{align}\label{j1}
-\mu^{2}(f+r^{2})-\frac{1}{A_{5}}\Big(-a^{2}J_{z}^{2}A_{2}^{2}+2aEJ_{z}A_{0}-E^{2}A_{1}^{2}+A_{5}^{2}\Big(\frac{\partial R}{\partial r}\Big)^{2}\Big) = C\,, \\
a^{2}E^{2}\sin^{2}\theta+\mu^{2}a^{2}\cos^{2}\theta-2aEJ_{z}+J_{z}^{2}\csc^{2}\theta+\Big(\frac{\partial \Theta}{\partial \theta}\Big)^{2} = C\,, \label{j2}
\end{align}
where $f$ is the function of $r$ represented in Eq. (\ref{dfgn}). $C$ appears as a separation constant and related to a more conventional Carter constant ($\mathcal{Q}$) as $\mathcal{Q}\equiv C-(J_{z}-aE)^{2}$. The functions $R(r)$ and $\Theta(\theta)$ can further be related to the respective momenta of the particle,
\begin{align}\label{j3}
\frac{dR}{dr} = \frac{\mu\Sigma}{A_{5}}\frac{dr}{d\tau} \hspace{7mm} ; \hspace{7mm} \frac{d\Theta}{d\theta} = \mu \Sigma \frac{d\theta}{d\tau}\,.
\end{align}
Thus, the expressions (\ref{j1}, \ref{j2}) take the following form
\begin{equation}
\begin{aligned}\label{dfvf}
\mu^{2}\Big(\frac{dr}{d\tau}\Big)^{2} =& \frac{A_{5}}{\Sigma^{2}}\Big[-C-\mu^{2}(f+r^{2})-\frac{1}{A_{5}}(-a^{2}J_{z}^{2}A_{2}^{2}+2aEJ_{z}A_{0}-E^{2}A_{1}^{2}) \Big]\,, \\
\mu^{2}\Big(\frac{d\theta}{d\tau}\Big)^{2} =& \frac{1}{\Sigma^{2}}\Big[C-(a^{2}E^{2}\sin^{2}\theta+\mu^{2}a^{2}\cos^{2}\theta-2aEJ_{z}+J_{z}^{2}\csc^{2}\theta) \Big]\,.
\end{aligned}
\end{equation}
Further, corresponding to two conserved quantities - energy and angular momentum - we obtain the following two relations,

\begin{equation}
\begin{aligned}\label{gdscs}
\mu\frac{dt}{d\tau}  =& \frac{1}{A_{5} \Sigma}(A_{1}^2 E+a J_{z} (A_{5}-A_{0}))+\mathcal{O}(a^{2})\,, \\
\mu\frac{d\phi}{d\tau} =& \frac{1}{A_{5} \Sigma}(A_{5} J_{z} \csc ^2\theta-a \left(A_{5}-A_{0}\right)E)+\mathcal{O}(a^{2})\,,
\end{aligned}
\end{equation}
% Finally, under the slow-rotation approximation, the geodesic velocities become
% \begin{align}\label{gdscs2}
% \mu\frac{dt}{d\tau}  =& \frac{1}{A_{5} \Sigma}(A_{1}^2 E+a J_{z} (A_{5}-A_{0})) \\ \label{gdscs3}
% \mu \frac{d\phi}{d\tau} =& \frac{1}{A_{5} \Sigma}(A_{5} J_{z} \csc ^2\theta-a \left(A_{5}-A_{0}\right)E) \\ \label{gdscs4}
% \mu^{2}\Big(\frac{dr}{d\tau}\Big)^{2} =& \frac{A_{5}}{\Sigma^{2}}\Big[-C-\mu^{2}(f+r^{2})-\frac{1}{A_{5}}(-a^{2}J_{z}^{2}A_{2}^{2}+2aEJ_{z}A_{0}-E^{2}A_{1}^{2}) \Big], \\
% \mu^{2}\Big(\frac{d\theta}{d\tau}\Big)^{2} =& \frac{1}{\Sigma^{2}}\Big[C-(a^{2}E^{2}\sin^{2}\theta+\mu^{2}a^{2}\cos^{2}\theta-2aEJ_{z}+J_{z}^{2}\csc^{2}\theta) \Big].
% \end{align}
% \begin{align}\label{gdscs2}
% \mu \frac{dt}{d\tau} =& \Big[-(r^{2}A_{1}A_{2}-\Delta)\frac{aJ_{z}}{r^{2}\Delta}+\frac{Er^{2}}{\Delta}A_{1}^{2}\Big]\Big(1+\epsilon_{3}\frac{M^{3}}{r^{3}}\Big)^{-1} + \mathcal{O}(a^{2})\,,\\ \label{gdscs3}
% \mu \frac{d\phi}{d\tau} =& \Big[\frac{J_{z}}{r^{2}\sin^{2}\theta}+\frac{aE}{r^{2}\Delta}(A_{1}A_{2}r^{2}-\Delta)\Big]\Big(1+\epsilon_{3}\frac{M^{3}}{r^{3}}\Big)^{-1}+ \mathcal{O}(a^{2})\,, \\ \label{gdscs4}
% \mu^{2}\Big(\frac{dr}{d\tau}\Big)^{2} =& \frac{A_{5}}{\Sigma^{2}}\Big[(A_{1}r^{2}E-aA_{2}J_{z})^{2}-\Delta\Big(Q+\mu^{2}r^{2}+\mu^{2}\epsilon_{3}\frac{M^{3}}{r}\Big)\Big] + \mathcal{O}(a^{2})\,, \\ \label{gdscs5}
% \mu^{2}\Big(\frac{d\theta}{d\tau}\Big)^{2} =& \frac{1}{\Sigma^{2}}\Big[Q-\Big(aE\sin\theta-\frac{J_{z}}{\sin\theta}\Big)^{2}\Big] + \mathcal{O}(a^{2}).
% \end{align}
under the slow-rotation approximation, $\Sigma=r^{2}+f(r)$. Using the four velocities, 
\begin{align}
\begin{split}
& \mu^{2}\Big[\Big(\frac{dr}{d\tau}\Big)^{2}+r^{2}\Big(\frac{d\theta}{d\tau}\Big)^{2}+r^{2}\sin^{2}\theta\Big(\frac{d\phi}{d\tau}\Big)^{2} \Big] = %A_{5}A_{1}^{2}E^{2}-\mu^{2}+\frac{2M\mu^{2}}{r}+\frac{2 A_{5} \mu ^2 M^4 \epsilon -4 \mu ^2 M^4 \epsilon}{r^4} \nonumber \\
%& \frac{-A_{5} M^3 \epsilon \Big(2 A_{1}^2 E^2+\mu ^2\Big)+2 A_{5} M Q+2 \mu ^2 M^3 \epsilon}{r^3} \nonumber \\ 
(E^2-\mu ^2)-\frac{M^2}{r^4} \left(-8 a E J_{z}+C (\alpha_{52}+4 \epsilon_{1})+2 \mu ^2 M^2 \epsilon_{3}\right)+ \\
& \frac{M}{r} \left(\mu ^2 (-\alpha_{51}+\epsilon_{1}+2)+2 E^2 (\alpha_{11}-\epsilon_{1})\right)+ \frac{M^2}{r^2} \left(\mu ^2 (-\alpha_{52}-2 \epsilon_{1}+\epsilon_{2})+2 E^2 (\alpha_{12}-\epsilon_{2})\right) \\
& +\frac{M}{r^{3}} \left(4 a E J_{z}-(\alpha_{51}-2) C+M^2 \left(2 E^2 (\alpha_{13}-\epsilon_{3})+\mu ^2 (\epsilon_{3}-2 \epsilon_{2})\right)\right)\,.
\end{split}
\end{align}
The above expression with respect to coordinate time can be written in the following form:

\begin{align}
\begin{split}
& \mu^{2}\Big[\Big(\frac{dr}{dt}\Big)^{2}+r^{2}\Big(\frac{d\theta}{dt}\Big)^{2}+r^{2}\sin^{2}\theta\Big(\frac{d\phi}{dt}\Big)^{2} \Big] =  \frac{E^2 \mu ^2-\mu ^4}{E^2}-\frac{1}{E^2 r}\Big(\mu^{4}M (-4 \alpha_{11}+3 \alpha_{51}+\epsilon_{1}-6)+  \\ 
& 2 E^2 \mu ^2 M (\alpha_{11}-\alpha_{51}+2)\Big) +  \frac{1}{E^{2}r^{2}}\Big(2 E^2 \mu ^2 M^2 (4 \alpha_{11}-\alpha_{12}-2 \alpha_{51}+\alpha_{52}+2)-  \\
& \mu ^4 M^2 (24 \alpha_{11}-4 \alpha_{12}-12 \alpha_{51}+3 \alpha_{52}-6 \epsilon_{1}+\epsilon_{2}+12)\Big)\,,
\end{split}
\end{align}
We have presented here the higher-order expansion up to $1/r^{2}$. However, the leading order emergence of deviations that contribute to the analysis appears at zero PN in $\mathcal{O}(1/r)$ itself, as can also be seen in final results mentioned in the main text Eqs. (\ref{avflxn}, \ref{dpdtn}, \ref{phase}). Further, we separate out the rest-mass energy by replacing $E=\mu+\mathcal{E}$ \cite{Misner:1973prb, Ryan:1995xi, Flanagan:2007tv} and keep terms only up to linear order in $\mathcal{E}$. Note that we ignore terms involving $\mathcal{O}(\mathcal{E} M/r)$ as well as their higher powers. Following \cite{AbhishekChowdhuri:2023gvu},
\begin{align}\label{ener2}
\mathcal{E} =& \frac{\mu}{2}\Big[\Big(\frac{dr}{dt}\Big)^{2}+r^{2}\Big(\frac{d\theta}{dt}\Big)^{2}+r^{2}\sin^{2}\theta\Big(\frac{d\phi}{dt}\Big)^{2} \Big]-\frac{\mu M}{r}+\frac{\mu  M }{2r}(-2 \alpha_{11}+\alpha_{51}+\epsilon_{1})  \nonumber \\
& +\frac{\mu  M^2}{2r^{2}} (16 \alpha_{11}-2 \alpha_{12}-8 \alpha_{51}+\alpha_{52}-6 \epsilon_{1}+\epsilon_{2}+8)\,.
\end{align}
With the linear order correction $\mathcal{E}$, the four-velocities are given as
\begin{equation}
\begin{aligned}\label{gdscs3n}
\Big(\frac{dr}{d\tau}\Big)^{2} = & 2 \mathcal{E}+\frac{1}{r}(2+2 \alpha_{11}-\alpha_{51}-\epsilon_{1})+\frac{1}{r^{2}}(-J_{z}^{2}+2 \alpha_{12}-\alpha_{52}-2 \epsilon_{1}-\epsilon_{2})+ \\
& \frac{1}{r^{3}}(2J_{z}^{2}-4 a J_{z}+2 \alpha_{13}-2 \epsilon_{2}-\epsilon_{3}+2 J_{z}^2 \epsilon_{1}-\alpha_{51} J_{z}^2)-\frac{1}{r^{4}}(J_{z}^2 (\alpha_{52}+4 \epsilon_{1}-2 \epsilon_{2})+2 \epsilon_{3})+ \\
& \frac{2 J_{z}^2}{r^5} (\epsilon_{3}-2 \epsilon_{2})-\frac{4 J_{z}^2 \epsilon_{3}}{r^6}\,, \\
\frac{d\phi}{d\tau} =& \frac{J_{z}}{r^2}+2a\Big(\frac{4}{r^5}+\frac{2}{r^4}+\frac{1}{r^3}\Big)-J_{z}\Big(\frac{\epsilon_{3}}{r^5}+\frac{\epsilon_{2}}{r^4}+\frac{\epsilon_{1}}{r^3}\Big)\,.
\end{aligned}
\end{equation}
We use these velocities to numerically estimate the rate change of orbital energy and angular momentum in the main text. Further, we use Eqs. (\ref{dfvf}, \ref{gdscs}) and compute, 

%\begin{equation}
\begin{align}
\mu^{2}r^{4}\Big[\Big(\frac{d\theta}{d\tau}\Big)^{2}+\sin^{2}\theta\Big(\frac{d\phi}{d\tau}\Big)^{2} \Big] =& C+2 a E J_{z}+\frac{4 a E J_{z} M-2 C M \epsilon_{1}}{r}+\frac{8 a E J_{z} M^2-2 C M^2 \epsilon_{2}}{r^2}+ \nonumber \\
& \frac{16 a E J_{z} M^3-2 C M^3 \epsilon_{3}}{r^3}\,.
\end{align}
%\end{equation}
Since we are considering the leading order PN corrections; hence, we may write down the same expression with respect to coordinate time. Further, using the angular component of the velocities,
\begin{align}\label{dphidt}
J_{z} = \frac{\mu\sin^2\theta }{r}\left(M^3 \epsilon_{3}+M^2 r \epsilon_{2}+M r^2 \epsilon_{1}+r^3\right)\frac{d\phi}{d\tau}-\frac{2 a E M \sin^2\theta }{r^{2}}(2 M+r).
\end{align}
The constants of motion are consistent with \cite{Flanagan:2007tv, Ryan:1995xi, AbhishekChowdhuri:2023gvu}. 
%The Eq.(\ref{ef}) to the $\mathcal{O}(a)$ takes the following form,
% \begin{align}
% \mu^{2}r^{4}\Big[\Big(\frac{d\theta}{dt}\Big)^{2}+\sin^{2}\theta\Big(\frac{d\phi}{dt}\Big)^{2} \Big] = \mathcal{Q}+J_{z}^{2}+a\Big(\frac{4EM}{r}+\frac{2EM^{2}\alpha_{22}}{r^{2}}\Big) \Big(1+\epsilon_{3}\frac{M^{3}}{r^{3}}\Big)\mu r^{2}\sin^{2}\theta \frac{d\phi}{d\tau},
% \end{align}
% or in the linear order corrections
% \begin{align}\label{cart1}
% \mathcal{Q}+J_{z}^{2} = \mu^{2}r^{4}(\Dot{\theta}^{2}+\sin^{2}\theta\Dot{\phi}^{2})-4a\mu^{2} Mr\Dot{\phi}\sin^{2}\theta.
% \end{align}
We restate that our analysis considers equatorial orbits. We use the velocities mentioned in this appendix and constants of motion to compute fluxes. We also add these expressions in the main body of the text. Note that, at this stage, we have kept terms involving various deformation parameters. However, there are only three parameters ($\alpha_{11}, \alpha_{51}, \epsilon_{1}$) contributing to the ultimate results obtained with leading order PN analysis. We indicate this point in the main body wherever it is needed.
%%%%%%%%%%%%%%%%%%%%%%%%%%%%%%%%%%%%%%%%%%%%%%%%%%%%%%%%%%%%%%%%%%%%%%%%%%%%%%%%%%%%%%%%%%%%%
\newpage
\section{Waveform}\label{apendwave}
\textcolor{black}{Here, we provide additional waveforms with distinct deviation parameters and eccentricities, comparing them to the GR part of the signal depicted in dotted curves.}
\begin{figure}[h!]
\centering

	\minipage{0.49\textwidth}
	\includegraphics[width=\linewidth]{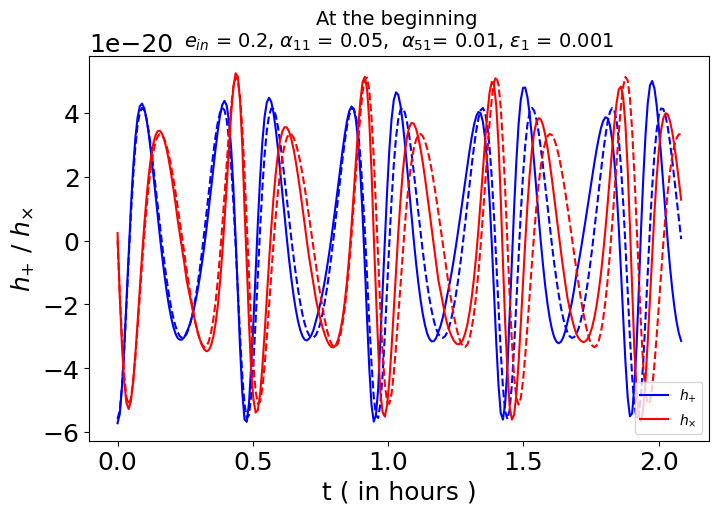}
% \caption{Wormholes for $\Lambda=0$}
	\endminipage\hfill
	%%%%%%%%%%%%%%%%%%%%%%%%
	\minipage{0.49\textwidth}
	\includegraphics[width=\linewidth]{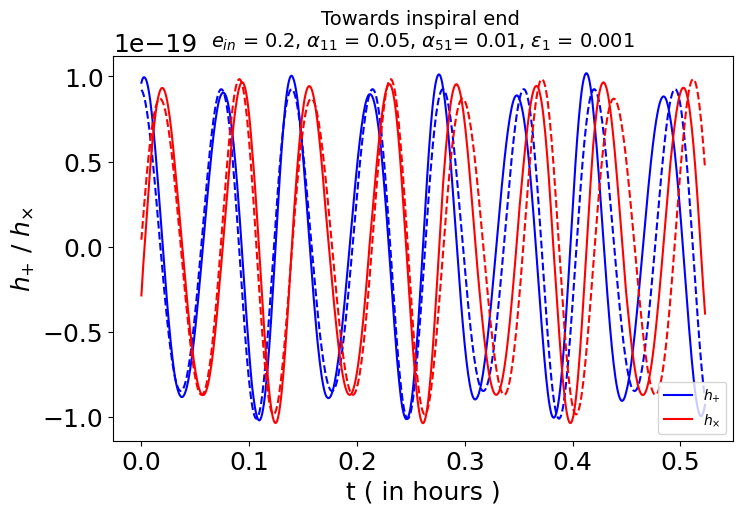}
	\endminipage\hfill
  \minipage{0.49\textwidth}
         \includegraphics[width=\linewidth]{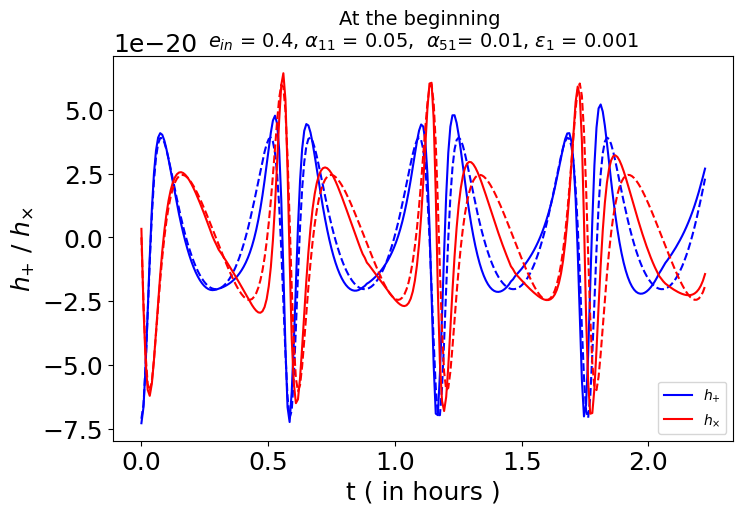}
	\endminipage\hfill
  \minipage{0.49\textwidth}
	\includegraphics[width=\linewidth]{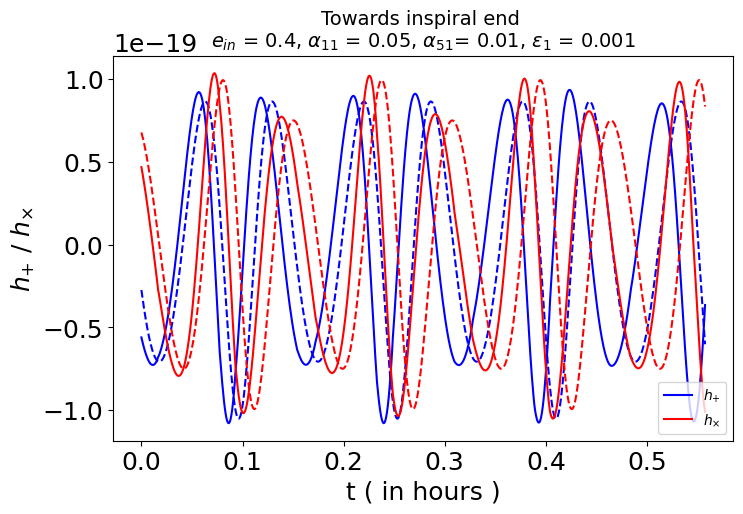}
	\endminipage
 
	\caption{The waveforms represent the GW signal with $\mu=10 M_\odot$ inspiralling the primary $M=10^{6}M_\odot$. The inspiral starts at $p=14$ and luminosity distance $R=10 \textup{Mpc}$. The non-zero deviations have been chosen for all three deviation parameters at $e_{\textup{in}}=0.2$ and $e_{\textup{in}}=0.4$. The left column shows the effects of the deviation parameters on the waveforms ($h_{+}, h_{\times}$) for the initial hours. The right column describes the same for the last few minutes towards the end of the inspiral. The dotted curves depict the GR part of the corresponding signal.} \label{waveform2}

\end{figure}
%%%%%%%%%%%%%%%%%%%%%%%%%%%%%%%%%%%%%%%%%%%%%%%%%%%%%%%%%%%%%%%%%%%%%%%%%%%%%%%%%%%%%%%%%%%%%%%%

% \section{Orbital phase}
% Here, we present some additional plots of the orbital phase that represent the deviations from the Kerr black hole for different values of ($\epsilon_{3}, \alpha_{52}$). 
% \begin{figure}[h!]
% 	%%%%%%%%%%%%%%%%%%%%%%%%
% 	\centering
% 	\minipage{0.48\textwidth}
% 	\includegraphics[width=\linewidth]{phase_append_e_0.3_a_0.05_epsilon3_varies.pdf}
% % \caption{Wormholes for $\Lambda=0$}
% 	\endminipage\hfill
% 	%%%%%%%%%%%%%%%%%%%%%%%%
% 	\minipage{0.48\textwidth}
% 	\includegraphics[width=\linewidth]{phase_append_e_0.5_a_0.05_alphs52_varies.pdf}
% 	\endminipage
% 	\caption{The behaviour of orbital phase for distinct values of ($\epsilon_{3}, \alpha_{52}$) is shown. The left panel shows the deviations with $e_{\textup{in}}=0.3$ and the right panel shows the same with $e_{\textup{in}}=0.5$. The dashed curve represents the Kerr result.
%  }\label{fig_enr_flux_same_einiAP}
% \end{figure}

%\newpage
%%%%%%%%%%%%%%%%%%%%%%%%%%%%%%%%%%%%%%%%%%%%%%%%%%%%%%%%%%%%%%%%%%%%%%%%%%%%%%%%%%%%%%%%%%%%%%%%%%%
\iffalse
\providecommand{\href}[2]{#2}\begingroup\raggedBig\endgroup

\fi

\bibliography{JN1}
\bibliographystyle{JHEP}
\end{document}